\documentclass[numberedappendix]{emulateapj}

\usepackage{graphicx, natbib}

\newcommand{\beq}	{\begin{equation}}
\newcommand{\eeq}	{\end{equation}}
\newcommand{\beqa}{\begin{eqnarray}}
\newcommand{\eeqa}{\end{eqnarray}}
\newcommand{\beqs}	{\begin{displaymath}}
\newcommand{\eeqs}	{\end{displaymath}}
\newcommand{\beqas}	{\begin{eqnarray*}}
\newcommand{\eeqas}	{\end{eqnarray*}}

\def\bit{\begin{itemize}}
\def\eit{\end{itemize}}

\def\simlt{\lower.5ex\hbox{$\; \buildrel < \over \sim \;$}}
\def\simgt{\lower.5ex\hbox{$\; \buildrel > \over \sim \;$}}
\def\la{\simlt}

\font\tenbi=cmmib10 
\newfam\bifam \def\bi{\fam\bifam\tenbi} \textfont\bifam=\tenbi
\font\tenbr=cmbx10
\newfam\brfam  \textfont\brfam=\tenbr

\font\squinttenbi=cmbx10 at 9pt
\scriptfont\brfam=\squinttenbi

\def\vecnabla{
              \setbox1=\hbox{$\bigtriangledown$}
                           \raise.45ex\hbox{$\bigtriangledown$\hskip-.97\wd1
                           $\bigtriangledown$\hskip-.97\wd1
                           $\bigtriangledown$\hskip-.97\wd1}
                           \raise.47ex\hbox{$\bigtriangledown$}}

\def\vecp{{\bi{p}}}

\def\vecv{{\bi{v}}}

\def\vecx{{\bi{x}}}

\def\vecB{{\bi{B}}}

\def\vecL{{\bi{L}}}

\def\symbol#1{\ifmmode#1\else$#1$\fi}

\newcommand{\jth}		{J_{\rm th}}
\newcommand{\jmax}	{J_{\rm max}}
\newcommand{\jthmax}	{J_{\rm th,\,max}}
\newcommand{\lj}		{\lambda_{\rm J}}
\newcommand{\mbe}	{M_{\rm BE}}
\newcommand{\mcr}	{M_{\rm cr}}

\newcommand{\rcr}		{R_{\rm cr}}
\newcommand{\rcrth}		{R_{\rm cr,\, th}}
\newcommand{\msun}{M_{\odot}}

\newcommand{\Orion}{\textsc{Orion}~}

\begin{document}

\title{The Fragmentation of Magnetized, Massive Star-Forming Cores with Radiative Feedback}

%\centerline{DRAFT: \today}
%\slugcomment{Accepted to the Astrophysical Journal}

\shorttitle{Fragmentation in Magnetized, Massive Star-Forming Cores}
\shortauthors{Myers et al.}

\author{
        Andrew T. Myers \altaffilmark{1},
        Christopher F. McKee\altaffilmark{1,2},
        Andrew J. Cunningham\altaffilmark{3},
        Richard I. Klein\altaffilmark{2, 3}, and
        Mark R. Krumholz\altaffilmark{4}      
        }

\altaffiltext{1}{Department of Physics, University of California, Berkeley,
Berkeley, CA 94720; atmyers@berkeley.edu}
\altaffiltext{2}{Department of Astronomy, University of California, Berkeley,
Berkeley, CA 94720}
\altaffiltext{3}{Lawrence Livermore National Laboratory, P.O. Box 808, L-23, Livermore, CA 94550}
\altaffiltext{4}{Department of Astronomy and Astrophysics,
         University of California, Santa Cruz, CA 95064}
         
\begin{abstract}
We present a set of 3-dimensional, radiation-magnetohydrodynamic calculations of the gravitational collapse of massive (300 $M_{\odot}$), star-forming molecular cloud cores. We show that the combined effects of magnetic fields and radiative feedback strongly suppress core fragmentation, leading to the production of single star systems rather than small clusters. We find that the two processes are efficient at suppressing fragmentation in different regimes, with the feedback most effective in the dense, central region and the magnetic field most effective in more diffuse, outer regions. Thus, the combination of the two is much more effective at suppressing fragmentation than either one considered in isolation. Our work suggests that typical massive cores, which have mass-to-flux ratios of about 2 relative to critical, likely form a single star system, but that cores with weaker fields may form a small star cluster. This result helps us understand why the observed relationship between the core mass function and the stellar initial mass function holds even for $\sim 100 M_{\odot}$ cores with many thermal Jeans masses of material. We also demonstrate that a $\sim 40$ AU Keplerian disk is able to form in our simulations, despite the braking effect caused by the strong magnetic field.
\end{abstract}


\section{Introduction}

Massive stars, which have mass  $> 8 M_{\odot}$, make up $< 1\%$ of the total stellar population, but their numbers belie their impact. Both the total luminosity and the ionizing luminosity of a star are highly super-linear functions of mass. Thus, massive stars have a much stronger impact on their birth environments than low-mass stars do. Since most stars form in clusters that contain at least one early O star, massive stars have an important impact on the formation of their low-mass neighbors, whether by altering the thermal properties of their parent clumps by heating the dust, or by destroying them outright via photoionization. The latter process is so bright that it allows observation of the star formation rate in other galaxies. Finally, massive stars end their lives in supernova explosions, which produce heavy elements and add large amounts of energy to the interstellar medium (ISM), contributing to the driving of its turbulence on large scales. Understanding the life cycle of massive stars from their births to their deaths is thus an important problem for many branches of astrophysics.

Unfortunately, the first stage of this process - the birth of massive stars - remains an incompletely understood problem. Observationally, regions of massive star formation in our own galaxy tend to lie farther away from Earth than regions of low-mass star formation, meaning that observers have not yet been able to probe the formation process for high-mass stars at the same level of detail as they have for low-mass stars. Theoretically, the central difficulty is the large number of mutually interacting physical processes involved. Massive stars form out of a supersonically turbulent, self-gravitating fluid with dynamically significant magnetic fields. Massive protostars also deeply impact their surroundings as they form through a variety of feedback processes, including magnetically-launched outflows, radiation pressure, radiative heating, and ionization. Because of the complexity of these processes, simulations of massive star formation are able to include at most a few of these effects at one time. In the past several years, there has been much work done on massive star formation that ignored the effects of magnetic fields, both with \citep[e.g.][]{krumholz2007b, krumholz2010, krumholz2009, cunningham2011} and without \citep[e.g.][]{girichidis2011} radiative feedback. There has also been much work on simulating massive star formation that included the magnetic field, but did not include radiative feedback \citep[e.g.][]{seifried2011, Seifried2012, li2006, wang2010, hennebelle2011}. Thus far only two published simulations of massive star formation have included both radiation and magnetic fields, and these provide only a limited picture of how fragmentation in massive cores works. \cite{peters2011} treat direct stellar radiation and ionization chemistry, but neglect the dust-reprocessed radiation field, which is mainly responsible for regulating fragmentation. \cite{commercon2011} include dust-reprocessed light, but because they do not employ a subgrid stellar model they are forced to halt their calculations when $\lesssim 1\%$ of the core material has collapsed, and as a result they cannot study the fragmentation of the bulk of the gas. 

In this paper, we attempt to fill that gap. We present the results of 3-dimensional, adaptive mesh refinement (AMR), radiation-magnetohydrodynamic (R-MHD) simulations that treat the dust-processed radiation from protostars in the flux-limited diffusion (FLD) approximation. In particular, we focus on the fragmentation of isolated, massive cores in the relatively early stages of star formation - up to the point at which about 10\% of the core gas has turned into stars. The question of how massive cores fragment is an important one for any theory of star formation in which the initial mass function (IMF) is set in the gas phase, e.g. the turbulent fragmentation scenario originally laid out in \cite{padoan2002}. Observations of the core mass function (CMF) in galactic star-forming regions reveal that it looks like a scaled-up version of the stellar initial mass function (IMF) \citep{alves2007, nutter2007, enoch2008}. This relationship appears to continue even up to $\sim 100 M_{\odot}$ \citep{reid2006}. This correspondence - that the CMF has the same form as the IMF but is shifted up in mass by a factor of $\sim 3$ - has a natural explanation if massive cores do not fragment strongly as they collapse, but instead simply convert $\sim 1/3$ of their mass into single massive stars or systems. 

The purpose of this paper is to address the question of how massive cores fragment via direct numerical simulation. Our outline is as follows: in section 2, we describe our numerical setup, including the equations and algorithms used as well as our initial and boundary conditions. In section 3, we present our results, focusing on the evolution of our cores over a period of 0.6 mean-density free-fall times. In section 4, we discuss our results, in which the magnetic field and the radiative transfer together have a significant impact on the fragmentation of the cores in a way one would not predict from either process considered in isolation. We summarize our conclusions in section 5. 

\section{Numerical Setup}
\subsection{Equations and Algorithms}
We solve the equations of mass, momentum, and energy conservation on a hierarchy of AMR grids. We assume that the motion of the gas is governed by the ideal MHD equations and treat the radiation using the mixed-frame approach of \cite{krumholz2007}. At any time, the computational domain consists of a fluid made up of gas, dust, and radiation, plus some number of sink particles that represent stars. The fluid quantities are described by a vector of state variables $(\rho, \rho \vecv, E, \vecB, E_R)$ defined at every grid cell, where $\rho$ is the gas density, $\rho \vecv$ the momentum, $E$ the non-gravitational energy density (i.e. the total of the kinetic, thermal, and magnetic energy densities), $\vecB$ the magnetic field, and $E_R$ the radiation energy density. The particles are characterized by their position $\vecx_i$, momentum $\vecp_i$, mass $M_i$, and luminosity $L_i$, which is determined via the protostellar evolution model described in \cite{mckee2003} and \cite{offner2009}. The equations governing the evolution of the R-MHD fluid-particle system are:

\begin{eqnarray}
\label{masscons}
\frac{\partial \rho}{\partial t} & = & - \nabla\cdot(\rho\vecv) - \sum_i \dot{M}_i W(\vecx-\vecx_i) \\
\frac{\partial(\rho \vecv)}{\partial t} & = & -\nabla\cdot(\rho \vecv\vecv - \frac{1}{4 \pi}\vecB\vecB) - \nabla P_T - \rho \nabla \phi - \lambda \nabla E_R
\nonumber \\
& & {} - \sum_i \dot{\vecp}_i W(\vecx-\vecx_i) 
\label{momcons}
\\
\frac{\partial E}{\partial t} & = & -\nabla \cdot [(E+P_T)\vecv -\frac{1}{4 \pi}\vecB(\vecv \cdot \vecB)] - \rho \vecv \cdot \nabla \phi 
\nonumber \\
& & {} - \kappa_{\rm 0P} \rho (4 \pi B_T - c E_R) + \lambda\left(2 \frac{\kappa_{\rm 0P}}{\kappa_{\rm 0R}} - 1\right) \vecv \cdot \nabla E_R 
\nonumber \\
& & {} - \sum_i \dot{\mathcal{E}}_i W(\vecx - \vecx_i) 
\label{econsgas}
\\
\frac{\partial\vecB}{\partial t} & = & - \nabla \cdot (\vecv\vecB - \vecB \vecv)
\label{induction}
\\
\frac{\partial E_R}{\partial t} & = & \nabla \cdot \left(\frac{c\lambda}{\kappa_{\rm 0R} \rho} \nabla E_R\right) + \kappa_{\rm 0P} \rho (4 \pi B_T - c E_R) 
\nonumber \\
& & {} - \lambda \left(2\frac{\kappa_{\rm 0P}}{\kappa_{\rm 0R}} - 1\right) \vecv\cdot \nabla E_R - \nabla \cdot \left(\frac{3 - R_2}{2} \vecv E_R\right)
\nonumber \\
& & {}
 + \sum_i L_i W(\vecx - \vecx_i).
\label{econsrad}
\end{eqnarray}

In the above equations, the total pressure $P_T$ is $P_{\rm{gas}} + B^2/8\pi$, and we use an ideal equation of state, so that
\begin{equation}
\label{EOS}
P_{\rm{gas}} = \frac{\rho k_B T_g}{\mu m_{\rm H}} = (\gamma-1) \rho \epsilon,
\end{equation} where $k_B$ is the Boltzmann constant, $T_g$ the gas temperature, $\mu$ the mean molecular weight, $\gamma$ the ratio of specific heats, and $\epsilon$ the thermal energy per unit mass. We take $\mu = 2.33$ and $\gamma = 5/3$, appropriate for molecular gas of solar composition that is too cold to store energy in rotational degrees of freedom. The corresponding value for the gas's specific heat capacity is $c_v = k_B / (\gamma - 1) \mu m_{\rm H} \approx 5.3 \times 10^7 $ erg g$^{-1}$ K$^{-1}$. 

The summations in the gas-sink interaction terms are taken over all the particles in the domain, and $W(\vecx - \vecx_i)$ is a weighting kernel that distributes the transfer of mass, momentum, and energy over a radius of 4 fine-level cells around sink particle $i$. The values for $\dot{M}_i$, $\dot{\vecp}_i$, and $\dot{\mathcal{E}}_i$, or the rates of mass, momentum, and energy transfer between the sink particles and the fluid, are computed by fitting the flow around each sink particle to a magnetized Bondi-Hoyle flow; see Lee et al. (2013, in preparation) for details. The star particle states themselves are updated according to the following equations:
\begin{eqnarray}
\label{starmass}
\frac{d}{dt} M_i &= & \dot{M}_i, \\
\label{starpos}
\frac{d}{dt} \vecx_i & = & \frac{\vecp_i}{M_i}, \\
\label{starmom}
\frac{d}{dt} \vecp_i & = & -M_i \nabla \phi + \dot{\vecp}_i.
\end{eqnarray} Because our sink particle algorithm destroys information about the fluid flow inside the 4 fine cell accretion zone around each particle, we are not able to properly follow the dynamics of particles that pass within that distance of each other. We therefore adopt the following criterion to handle mergers between sink particles that pass within one accretion radius of each other (40 AU in most of the simulations presented here):  we merge the two sinks together only if the smaller sink is less than $0.05 M_{\odot}$ in mass. This threshold roughly corresponds to the mass at which second collapse occurs \citep{masunaga98, masunaga2000}. Before that point, sink particles represent hydrostatic cores of several AU in size, which could be expected to merge together. After that point, they have collapsed down to roughly solar size scales, and will not necessarily merge simply because they pass within 40 AU of each other. 

The gravitational potential $\phi$ in the above expressions obeys the Poisson equation with a right-hand side that includes contributions from both the fluid and the star particles:
\begin{equation}
\label{poisson}
\nabla^2\phi = -4\pi G \left[ \rho + \sum_i M_i \delta(\vecx-\vecx_i)\right], 
\end{equation} where $G$ is the gravitational constant.

The radiation-specific quantities are the speed of light $c$, the comoving frame specific Planck- and Rosseland-mean opacities $\kappa_{\rm 0R}$ and $\kappa_{\rm 0P}$, and the Planck function $B_T = c a_R T_g^4 / (4\pi)$, where $a_R$ is the radiation constant. Finally, the flux limiter $\lambda$ and Eddington factor $R_2$ are two quantities that enter the flux-limited diffusion approximation we use to compute the radiative transfer. In this work, we adopt the \citet{levermore81} approximation:
\begin{eqnarray}
\lambda & = & \frac{1}{R} \left(\mbox{coth} R - \frac{1}{R}\right) \\
R & = & \frac{|\nabla E|}{\kappa_{\rm 0R} \rho E} \\
R_2 & = & \lambda + \lambda^2 R^2.
\end{eqnarray}
We obtain the dust opacities $\kappa_{\rm 0P}$ and $\kappa_{\rm 0R}$ from a piecewise-linear fit to the models of \citet{semenov2003}; see \citet{cunningham2011} for the exact functional form.

We solve the above equations using a new version of our astrophysical AMR code {\textsc{Orion}}, which allows us to simultaneously include the magnetic field and the radiative feedback. \Orion solves the above equations in a number of steps, which we summarize below. First, we solve the ideal MHD equations by themselves (Equation (\ref{induction}), the first two terms of Equation (\ref{masscons}) and the first three terms of Equations (\ref{momcons}) and (\ref{econsgas})) using a Godunov-type scheme with the HLLD approximate Riemann solver \citep{miyoshi2005}. Specifically, we use the dimensionally unsplit, AMR Constrained Transport (CT) scheme described in \cite{li2012}, which makes use of the unigrid CT scheme from the open-source astrophysical MHD code Pluto \citep{pluto}. This portion of the update algorithm uses a face-centered representation for the magnetic field $\vecB$, and we use the Chombo AMR library to provide support for the face-centered fields. Next, we incorporate self-gravity in the manner of \cite{truelove98} and \cite{klein99}. To solve the Poisson equation (Equation (\ref{poisson})), we use an iterative multigrid scheme also provided by Chombo. In the third step, we update Equations (\ref{momcons}), (\ref{econsgas}), and (\ref{econsrad}) for the radiative terms using the operator-split approach described in \cite{krumholz2007}. Briefly, this technique first solves the radiation pressure, work, and advection terms explicitly, and then implicitly updates the gas and radiation energy densities for the terms involving diffusion and the emission/absorption of radiation. This update is handled by the iterative process described in \cite{shestakov2005}, which uses psuedo-transient continuation to reduce the number of iterations required for convergence. We then complete the update cycle by calculating the new sink particle states using the above equations and computing their interactions with the fluid using the algorithms described in Lee et al. (2013, in preparation). 

Finally, we point out some important numerical caveats: our treatment of the radiation in this work focuses on the diffuse, dust-processed component of the radiation field, and it treats that radiation as gray. Massive stars, however, put out large numbers of ionizing photons, and these photons have a dramatic impact on the surrounding environment. Furthermore, treating the diffuse component of the field as gray and ignoring the direct component of the non-ionizing radiation both lead us to underestimate the radiation pressure force by a factor of a few \citep{kuiper2011}. However, since both of these effects are most significant for stars more massive than $\sim 20 M_{\odot}$, and since our conclusions are mainly based on the evolution of the cores prior to the most massive star reaching that point, we do not believe that our qualitative conclusions will be significantly altered by a more accurate treatment of the radiative transfer. We have also not included the effects of protostellar outflows in any of the runs in this paper. We shall do so in future work. 

\subsection{Refinement and Sink Creation}

The computational domain is a cube with side $L_{\text{box}}$ that is discretized into a coarse grid of $N_0$ cells, so that the resolution on the coarse grid $ \Delta x_0 = L_{\text{box}} / N_0$. Our code operates within an AMR framework that automatically adds and removes finer grids as the simulations evolve. With $L$ levels of refinement and a refinement ratio of 2, the resolution of the finest level is $\Delta x_L$ is $\Delta x_0 / 2^L$.  In this work, we have chosen these parameters such that $\Delta x_L$ is $10$ AU. 

Any cell that meets one or more of the following criteria is flagged for refinement:

\begin{enumerate}
\item The density in the cell exceeds the magnetic Jeans density, given by 
\begin{equation}
\rho_{\rm max}=\frac{\pi \jmax^2 c_s^2}{G\Delta x_l^2}\left(1+\frac{0.74}{\beta}\right).
\label{rhomax}
\end{equation}
where $c_s$ is the isothermal sound speed, $\Delta x_l$ the cell size on level $l$, $\beta = 8 \pi \rho c_s^2 / B^2$ and $\jmax$ is the maximum allowed number of magnetic Jeans lengths per cell, which must be small to avoid artificial fragmentation. Throughout this work, we take $\jmax = 1/8$. Note that this is identical to our previous work except for the inclusion of the magnetic field. Because the field provides additional support against collapse, we do not need to resolve the flow as highly in the presence of magnetic fields to prevent artificial fragmentation. For a derivation and numerical justification of this relation, see the Appendix, but we note that it is roughly equivalent to including the magnetic energy density along with the thermal energy in the expression for the Jeans length.  
\item The cell is within 16 $\Delta x_l$ of a sink particle.
\item The gradient in the radiation energy density exceeds 
\begin{equation}
\nabla E_R > 0.25 \frac{E_R}{\Delta x_l}.
\end{equation}
\end{enumerate} This procedure is repeated recursively until the final level is reached. At that point, if there are still any cells on the finest level that exceed the magnetic Jeans density, then the excess matter is removed from the cell and placed into a new sink particle, which then evolves according to the algorithm in section 2.1 above. Taken together, these three conditions ensure that the regions where star formation is happening are always tracked with the highest available numerical resolution. 

The application of these criteria to simulations of self-gravitating, isothermal gas requires special care, because such simulations have a fundamental problem: They do not converge. Isothermal gas tends to produce long, thin filaments, which do not fragment strongly \citep{inutsuka92, truelove98} and are thus non-trivial to decompose into point particles. Convergence studies by Boss et al. (2000) and Martel et al. (2006) suggest that there is no well-defined, converged solution for fragmentation and sink particle creation in this case, because the correct solution is collapse to singular filaments rather than singular points. As a result, for any choice of the finest resolution, application of the Truelove criterion to a collapsing isothermal gas will result in producing artificial fragments at the finest grid scale. This does not mean that all fragmentation in isothermal simulations is artificial: As we shall see below, our isothermal simulation produces about the same total mass in stars and the same amount of mass in the most massive star as our radiative simulations; on the other hand, it produces many more low-mass stars. In view of this over-fragmentation problem in isothermal simulations of star formation, it is essential to carry out a resolution study to verify that the conclusions being drawn from such simulations are physical and not numerical.

Interestingly enough, while much of the fragmentation in isothermal simulations is ultimately caused by the numerical mesh, proper adjustment of the finest level of resolution may nonetheless enable isothermal simulations to give a qualitatively correct picture of fragmentation in the absence of radiative feedback. Without protostellar heating, molecular gas still becomes non-isothermal at some density $\rho_{\rm{crit}}$ at which energy can no longer be efficiently radiated away. \cite{masunaga99} find that, for our choice of initial temperature and dust opacity, $ \rho_{\rm{crit}} \sim 10^{-13}$ g cm$^{-3}$. Past that point, the thermal pressure inside the filament starts to become more important relative to gravity. Eventually, gravitational contraction begins to slow, the timescale for cylindrical collapse becomes large compared to that for spherical collapse, and fragmentation will occur. Unfortunately, the results from such a simulation cannot be validated with a convergence study: increasing the resolution makes the fragments that form smaller than appropriate for the actual, non-isothermal case.

We stress that the non-convergence of the number of fragments in isothermal simulations is not a consequence of our particular sink particle algorithm. Using more stringent sink creation criteria, like those proposed in \cite{federrath2010}, has the benefit of producing fewer spurious fragments, but some will still be present, and their properties will still ultimately be determined by the numerical mesh. Furthermore, one cannot get around this problem by suppressing sink formation entirely within filamentary structures, since once the Truelove criterion is violated the filament will fragment artificially anyway. To get a converged answer on the number of fragments formed in self-gravitating, turbulent media, one must include some sort of deviation from isothermality and a fine enough numerical mesh to resolve the resulting fragments. 

\subsection{Initial and Boundary Conditions}

\begin{deluxetable*}{ccccccccccccc}
\tablecaption{Simulation Parameters\label{runsetup}}
\tablehead{
\colhead{Name} &
\colhead{RT?} &
\colhead{$M$ $(\msun)$} &
\colhead{$R$ (pc)} &
\colhead{$\sigma_v$ (km s$^{-1}$)} &
\colhead{$t_{\rm ff}$ (kyr)} &
\colhead{$M / M_\Phi$} &
\colhead{$\bar B$ (mG)} &
\colhead{$\bar \beta$} & 
\colhead{$L_{\rm box}$ (pc)} &
\colhead{$N_0$} &
\colhead{$L$} &
\colhead{$\Delta x_L$ (AU)}
}
\startdata        
HR& Yes  &300 & 0.1 & 2.3 & 30.2 & $\infty$ & 0.0 & $\infty$ & 0.4 & 256 & 5 &10.0  \\
BR & Yes  &300 & 0.1 & 2.3 & 30.2 & 2.0         & 1.6 & 0.05 & 0.4 & 256 & 5 &10.0  \\
BI  & No &300 & 0.1 & 2.3 & 30.2 & 2.0         & 1.6 & 0.05 & 0.4 & 256 & 5 &10.0  \\
\enddata
\tablecomments{Col.\ 8: mean magnetic field in the core. Col.\ 9: mean plasma $\beta = 8 \pi \rho c_s^2 / B^2$ in the core. Col.\ 10: resolution of the base grid. Col.\ 11: number of levels of refinement. Col.\ 12: maximum resolution at the finest level.
}
\end{deluxetable*}

We begin with three cores that are identical except that we include a different combination of physical processes in each run. The parameters for these simulations are summarized in Table \ref{runsetup}. Run HR includes the radiative transfer physics but has no magnetic field, run BI has a magnetic field but no radiation, and run BR has both a magnetic field and the radiative transfer. For run BI, we have dropped Equations (\ref{econsgas}) and (\ref{econsrad}) and adopted the isothermal equation of state ($P_{\rm{gas}} = \rho {c_s}^2$) instead of Equation (\ref{EOS}).

With the exception of the magnetic field, our initial conditions are almost identical to the those in \cite{myers2011} and (with the exception of the protostellar outflows) \cite{cunningham2011}. In all of our runs, we begin with an isolated sphere of gas and dust with mass $M_c = 300$  $\msun$, radius $R_c = 0.1 $ pc, and temperature $T_c = 20$ K. The density follows a power-law profile proportional to $r^{-1.5}$, so that the density at the edge of the core is 
\beq
\rho_{\rm{edge}} = \frac{3 M_c }{8 \pi R_c^3}.
\eeq The surface density of these cores, $\Sigma_c = M _c/ \pi R_c^2 \approx 2.0$ g cm$^{-2}$, is chosen to resemble that observed in galactic regions of high-mass star formation. For example, \cite{mckee2003} inferred a mean $\Sigma \sim 1$ g cm$^{-2}$ from the sample of high-mass clumps in \cite{plume1997}. The corresponding mean density is $\bar\rho \approx 4.8 \times 10^{-18}$ g cm$^{-3}$, or $\bar n_{\rm H} = 2.4 \times 10^{6}$ H nuclei cm$^{-3}$. This value determines the characteristic timescale for gravitational collapse, given by
\beq
t_{\rm{ff}} = \sqrt{\frac{3 \pi }{32 G \bar\rho}} \approx 30.2 \hspace{ 3 pt} \rm{kyr}.
\eeq While these initial parameters are to an extent chosen for computational convenience (higher densities mean shorter free-fall times, which mean fewer total time steps need to be taken) they are consistent with sub-mm interferometric observations of massive cores \citep{swift2009}. Furthermore, the $r^{-1.5}$ density profile agrees with observations of star-forming regions at the $\sim 1$ pc clump scale (\cite{beuther2007}, \cite{caselli95}, \cite{mueller2002}) and the $\sim 0.1$ pc core scale \citep{longmore2011, zhang2009}. Similarly, a recent mid-infrared extinction study \citep{butler2012} observed 42 massive cores in 10 different IRDCs and (after envelope subtraction) reported a mean $k_{\rho}$ of $\approx 1.6$. They also report that the power-law profile was a better fit to their observations than the less centrally concentrated Bonnor-Ebert profile. 

Our cores are placed at the center of a cubic box with side length equal to 0.4 pc, so that the sides are far enough removed from the core that there is minimal interaction from the boundaries. The parts of the box that are not covered by the core are filled with a hot, diffuse medium with $\rho_m = \rho_{\rm{edge}} / 10$ and $T_m = 200 $ K, so that the ambient medium will be in thermal pressure equilibrium with the core. We set the opacity of this confining gas to zero so that it will not cool as the simulation proceeds. The initial condition on $E_R$ is given everywhere by $a_R T_R^4$, where the radiation temperature $T_R$ is also set to 20 K. 

For boundary conditions, we choose outflow for the MHD update, meaning that in advancing the hyperbolic subsystem we set the gradients of $\rho, \rho \vecv, E,$ and $\vecB$ to zero at the domain boundary. For the radiation update, we use Marshak boundary conditions, meaning that the entire simulation volume is bathed in a blackbody radiative flux corresponding to 20 K, while radiation generated within the simulation volume may escape freely. Finally, in solving Equation (\ref{poisson}) for $\phi$, we require that $\phi = 0$ at the boundaries.  

We also give the core an initial 1D velocity dispersion of $\sigma_c = 2.3 $ km s$^{-1}$, chosen to put the core into approximate virial balance. If we take the virial ratio $\alpha$ to be $5 \sigma_c^2 R_c / G M_c$ \citep{bertoldi92}, then $\alpha \approx 2.1$. Thus, there is initially slightly more kinetic energy than gravitational potential energy in each of our cores.  We choose a slightly super-virial value for $\alpha$ because we do not drive the turbulence by adding kinetic energy after the simulations begin. Although the virial parameter greater than unity at $t = 0$, it has decayed to $\approx 1.0$ by the time the simulations end. The velocities themselves are drawn from a Gaussian random field with power spectrum $P(k) \propto k^{-2}$, appropriate for the highly supersonic turbulence found in molecular cloud cores. We include the perturbations in the following manner: first, we generate a $1024^3$ perturbation cube using the method of \cite{dubinski1995} with power on scales ranging from $k_{\rm{min}} = 1$ to $k_{\rm{max}} = 512$. We then place the cube over the simulation volume and either coarsen or interpolate the perturbation data so that we can represent perturbations at all levels of refinement. We have made no attempt to filter out compressive modes from the initial velocity field. The precise mixture of solenoidal and compressive components have been found to be important for gravitational fragmentation in unforced core collapse simulations \citep{girichidis2011} and on the overall rate of star formation in simulations with driven turbulence \citep{federrath2012}, but we do not explore this effect here. 

In our MHD runs, we also give the cores an initial magnetic field pointing in the \it z \rm direction. The importance of this field is best expressed in terms of the mass-to-flux ratio:
\beq
\mu_{\Phi} = M / M_\Phi, 
\eeq where
\beq
M_\Phi\simeq\frac{\Phi}{2\pi G^{1/2}}
\eeq is the magnetic critical mass and $\Phi$ is the magnetic flux threading the core. Cores with $\mu_{\Phi} > 1$ are unstable against gravitational collapse, while cores with $\mu_{\Phi} < 1$ are expected to be stable. Measurements of Zeeman splitting in both the OH molecule \citep{troland2008}, which probes densities of $10^{3-4}$ cm$^{-3}$, and the CN molecule \citep{falgarone2008}, which probes higher densities of $10^{5-6}$ cm$^{-3}$, show that the mean value of $\mu_{\Phi}$ is approximately 2, a value supported by theoretical arguments as well \citep{mckee1989}. Note, however, that there may be substantial scatter in the magnetic field strength such that many dark molecular cloud cores have much more supercritical values of the mass-to-flux ratio \citep{crutcher2010}. In this paper, we adopt $\mu_{\Phi} = 2$ for all of our MHD runs, and defer a more extensive parameter study on the effects of the magnetic field strength to a later work. 

In the absence of more detailed information about the magnetic field geometry, we will assume that the spatial dependence of the initial $\vecB$ field follows the cylindrically symmetric profile
\beq
\vecB(R_z) = B_{\rm{edge}} \left(\frac{R_z}{R_c}\right)^{-1/2} \hat z,
\eeq where $R_z$ is the distance to the z axis and the value of $B_{\rm{edge}}$ is chosen to give the desired mean mass-to-flux ratio for overall core:
\beq
B_{\rm{edge}} = \frac{3}{2} \frac{\sqrt{G} M_c}{\mu_{\Phi} R_c^2}. 
\eeq For $\mu_{\Phi} = 2$,  $B_{\rm{edge}} \approx 1.2$ mG. Using this form for the initial magnetic field is clearly an idealization, but it does have the advantage that it 1) satisfies the condition $\nabla \cdot \vecB = 0$, and 2) ensures that the mass-to-flux ratio in the central flux tube ($\sim 5.6$ above critical) does not greatly exceed the mean value for the overall core, consistent with the Zeeman measurements discussed above. 

Our initial conditions do not include any explicit rotation on top of the random turbulent perturbations described above. However, these perturbations do include some incidental angular momentum. In fact, as found by \cite{BB2000}, Gaussian random turbulence alone may be sufficient to account for the observed rotational properties of prestellar cores. When we apply the technique in that paper to measure $\beta_{\rm{rot}}$ for our cores, we get get $\beta_{\rm{rot}} = 0.012$, in line with the values observed in \cite{Goodman93}. Note, however, that as discussed in \cite{dib10}, the rotational properties of cores measured in projection by observers may differ substantially from the actual 3D values. In fact, if we calculate $E_{\rm{rot}} / E_{\rm{grav}}$ from our initial conditions using the full 3D velocity and density information, we get $ \approx 0.002$, lower than $\beta_{\rm{rot}}$ by a factor of 6. Thus, while the rotation in our initial conditions is consistent with observations, it is significantly lower than in other simulations that impose solid-body rotation in addition to random turbulence, such as those of \cite{Seifried2012}. Finally, as we do not chose the direction of the angular momentum vector in our cores explicitly, there was no imposed choice about the initial orientation of the core angular momentum vector $\vecL$ with respect to $\vecB$. It turns out to be misaligned with the magnetic field by $\theta \approx 60$ degrees.

While the above initial conditions are clearly somewhat artificial, they do capture the essential observed properties of high-mass dark-cloud cores. The most unrealistic aspect of our initial conditions is probably our imperfect treatment of the initial turbulence. While we include perturbations to the velocity field, there are no corresponding perturbations to the density at time $t = 0$. Thus, while the velocity field soon creates filamentary structures reminiscent of those expected from turbulence, these filaments do not have the same properties they would in a self-consistent realization of a turbulent density-velocity field, as discussed in \cite{krumholz2012} and \cite{federrath2012}. \cite{krumholz2012} found that this difference can have an important impact on e.g. the overall star formation rate, so we mention it here as a caveat. Another caveat is that our initial velocity field does not include any infall motions at $t = 0$. This probably has the effect of encouraging fragmentation somewhat, since the accretion rates and therefore the protostellar heating rates would be higher if infall were included from the beginning. Ideally, one would generate initial conditions for massive cores from larger simulations at the clump scale, which would then contain self-consistent density perturbations and infall. We are considering these issues in simulations of massive star formation at the cluster scale that are now in progress. The goal of this paper is to examine an idealized case first to elucidate the underlying physics.

We wish to emphasize that we have chosen the above runs to as far as possible create a controlled experiment where we have isolated the effect of only one physical process. Runs BR and BI are identical expect for the presence of the radiative feedback, and runs HR and BR are identical except for the presence of the magnetic field. Thus, we can isolate the effect of the radiative feedback by comparing the first set of runs, and the effect of the magnetic field by comparing the second. 
 
 \section{Results}
 
 Here, we summarize the main results of our calculations. The simulations presented here were run on the NASA supercomputing platform Pleiades on 128 to 512 processor cores and took a total of about 700,000 CPU hours. 
 
 \subsection{Density Structure}

 \begin{figure*}
  \epsscale{0.76}
  \centering
  \plotone{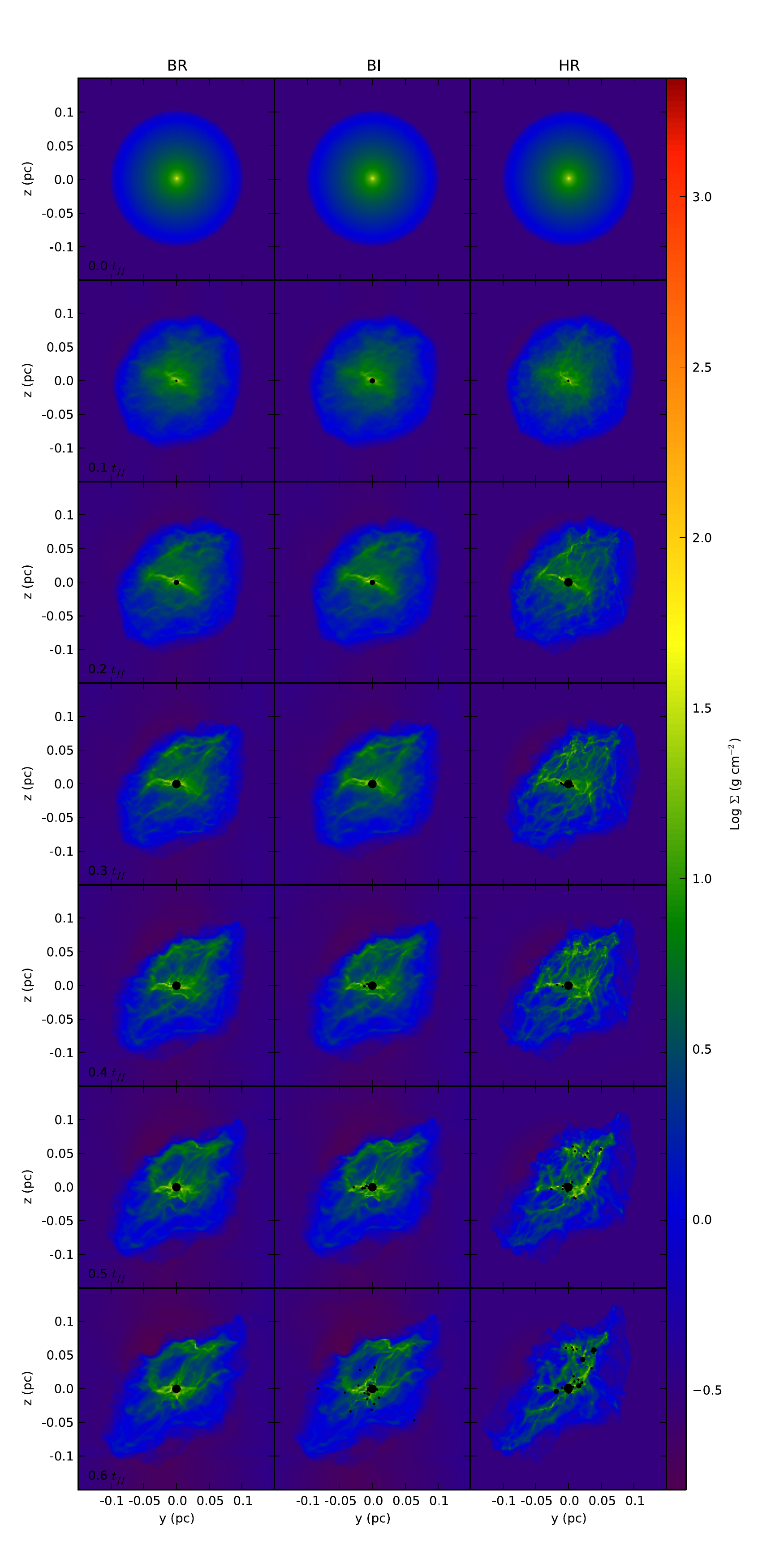}
  \caption{Column density through the simulation volume at 6 different times for runs BR (left), BI (middle), and HR (right).  Projections are taken along the \it x \rm direction, and the initial magnetic field is oriented in the positive \it z \rm direction. We have set the viewing area of the images to be 0.3 by 0.3 pc to show the global evolution of the entire core. Star particles are portrayed as black circles, with the size of the circle corresponding to the mass of the star. The smallest circles represent stars with masses between $0.05 M_{\odot}$ and $1.0 M_{\odot}$. The next size up represents masses between $1.0 M_{\odot}$ and $8.0 M_{\odot}$, and the largest represents stars with masses greater than $8.0 M_{\odot}$.}
  \label{fig:ColumnDensityZoomedOut}  
\end{figure*}

\begin{figure*}
  \epsscale{0.76}
  \centering
  \plotone{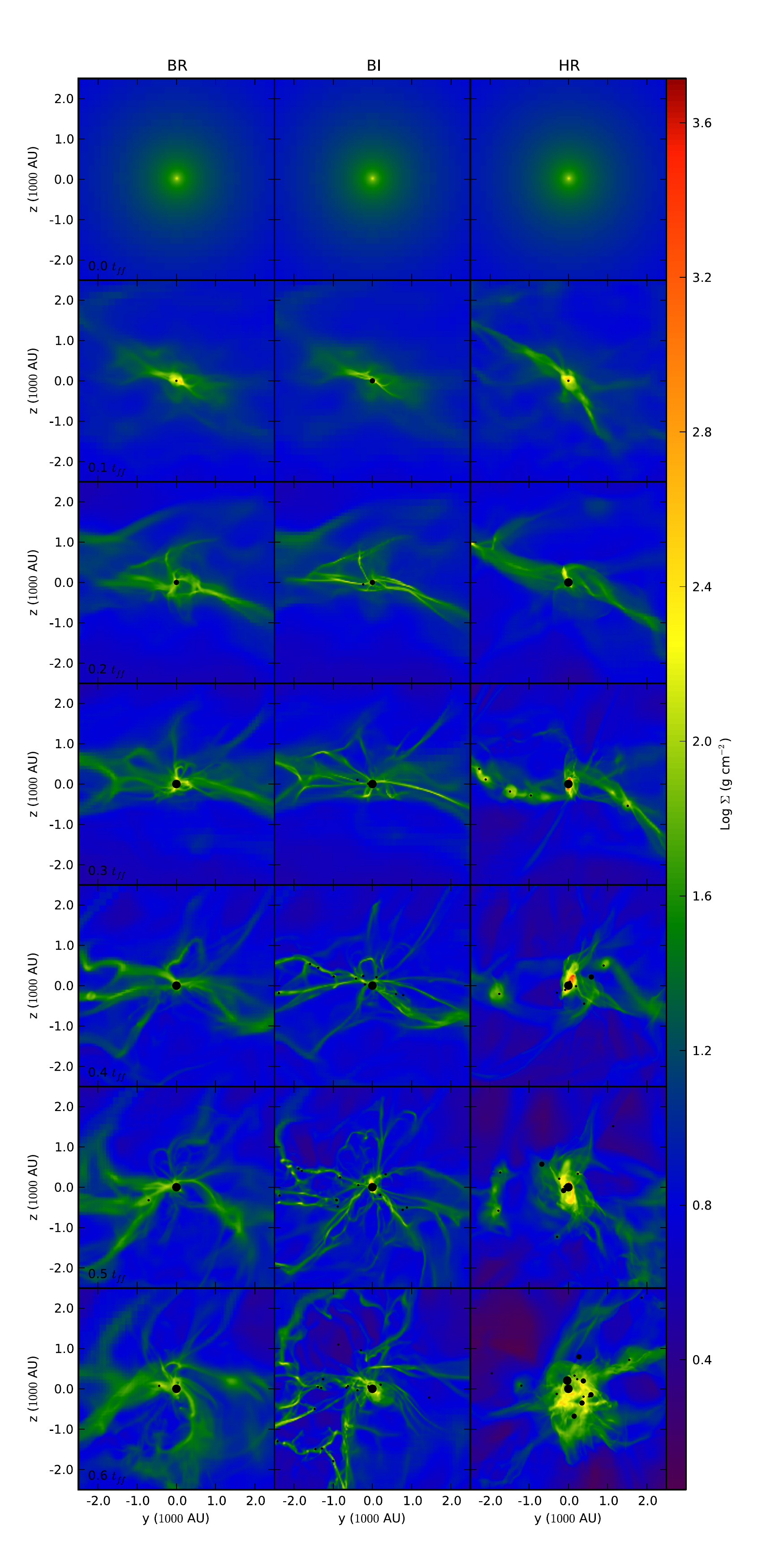}
  \caption{Same as Figure 1, but zoomed in to show the central 5000 AU around the most massive star in each simulation. Projections are still taken along the $x$ direction through the entire simulation volume.}
  \label{fig:ColumnDensityZoomedIn}
\end{figure*}
 
The time evolution of the large-scale structure of cores BR, HR, and BI is shown in Figure \ref{fig:ColumnDensityZoomedOut}. In all three runs, the imposed velocity perturbations create a system of filaments embedded within the collapsing core that feed gas into the central region where the massive star is forming. In the MHD runs the velocity perturbations rearrange the field lines so that the filaments are primarily perpendicular to the field. At this scale, the primary difference between the runs is that the filamentary structure created by the velocity perturbations in run HR is much more pronounced than in either of the runs with a magnetic field, despite the fact that all three runs have the same sonic Mach number of $\sim 15$. There are two reasons for this behavior. First, even though the cores in runs BR and BI are highly supersonic, they are only marginally super-Alfvenic, with ${\cal{M}_A} \approx 1.9 $. The presence of the faster magnetic signal speeds means that although shocks parallel to the magnetic field lines can be as strong as in run HR, flows perpendicular to the field that would be strong shocks in run HR are only weak shocks - or not shocks at all - in the other two runs. The overall effect is that, even ignoring gravity, the density contrasts imposed by the turbulence in the MHD runs are smaller than the hydro only run. Second, in all three runs, over-densities created by the turbulence can grow due to the self-gravity of the gas. However, in the presence of the magnetic field, these dense regions are only able to grow by drawing in material along the field lines, whereas there is no such restriction in the hydrodynamic case. The combined effect is that density distribution in the cores at a given time is broader in run HR than in the other two - that is, the dense regions are more dense and the diffuse regions more diffuse. Finally, we note in passing that at this scale the effect of the radiative heating has essentially no effect on the morphology of the core; the gas structure in runs BR and BI appears practically identical. 

The situation is different when we zoom in to show the central 5000 AU of the simulation volume as in Figure \ref{fig:ColumnDensityZoomedIn}, where the center is defined as the location of the most massive star in the simulation. At this scale, we begin to see clear differences in the gas morphology between runs BR and BI. In both cases, the gas collapses into a network of filaments, and there is a rough correspondence between the filaments in BR and those in BI. However, the filaments in run BR are much fatter and more diffuse than in run BI. This is easily understood as a consequence of radiative heating. For an isothermal, magnetized filament like the ones in BI, both the magnetic and pressure forces scale the same way with filament size as gravity in the virial theorem (see the Appendix for a more detailed discussion). Thus, either the total pressure (magnetic plus thermal) is initially enough to halt collapse, or else it will never be and the filament will collapse until something causes the equation of state to deviate from isothermality \citep{inutsuka2001}. This behavior is clearly seen in run BI, where the filaments contract until they reach the density at which our code creates sink particles. In run BR, on the other hand, radiative feedback from the central protostar has already caused the gas to become non-isothermal, and thus filaments close to the protostar stop collapsing before much sink creation takes place. 
 
\begin{figure}
  \epsscale{2.5}\plottwo{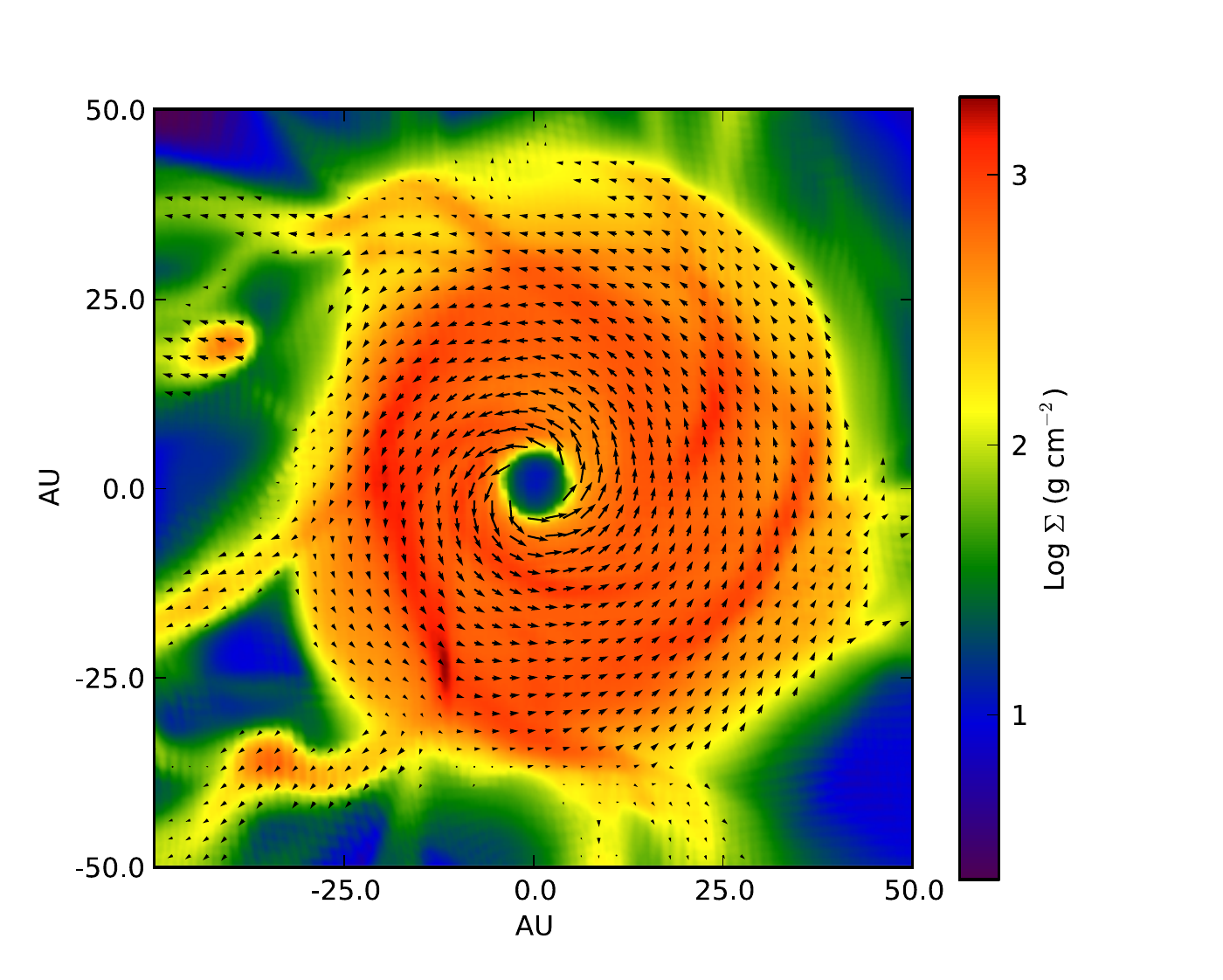}{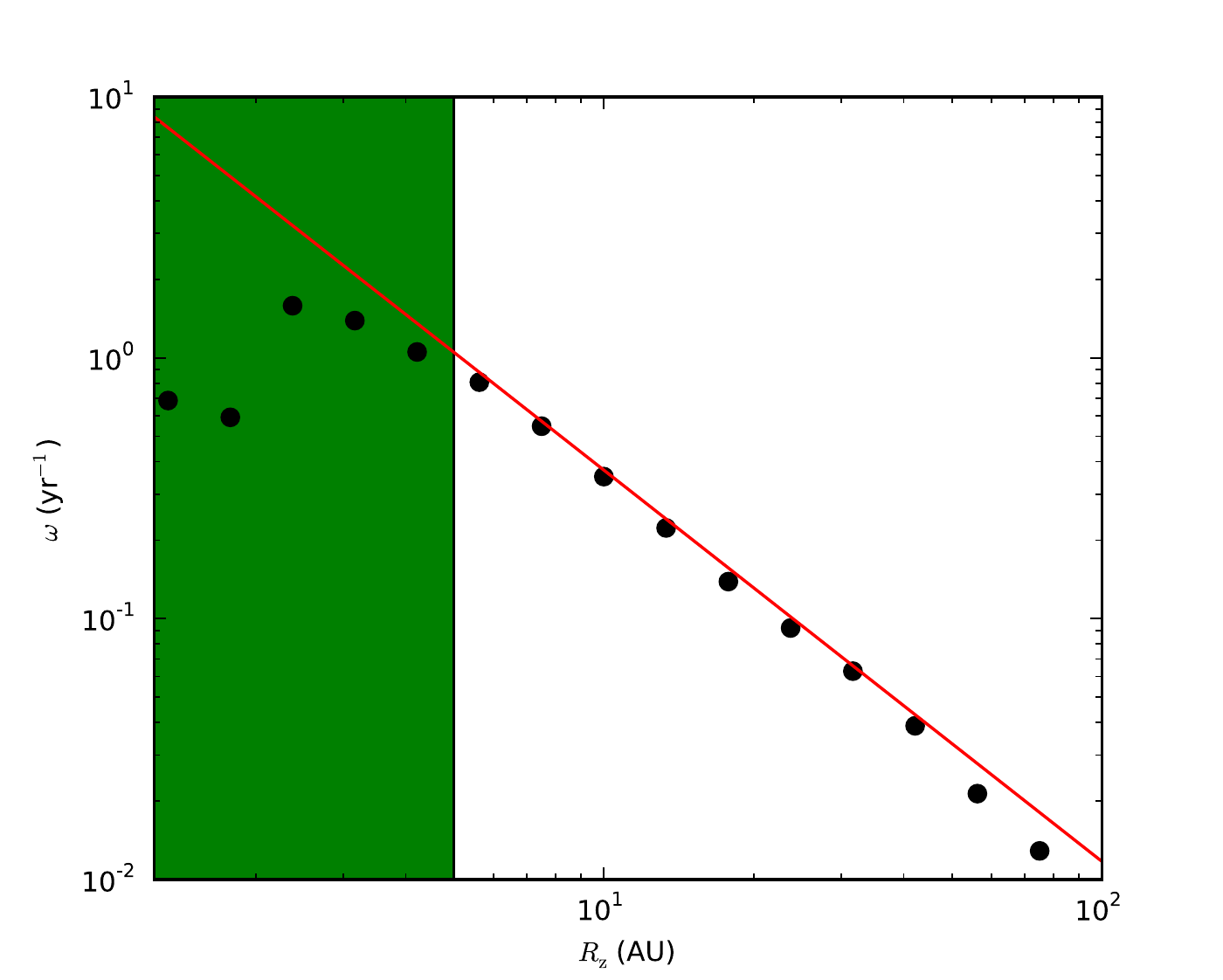}
  \caption{Top - Face-on view of the disk in the high-resolution version of run BI at 0.2 $t_{\rm{ff}}$. The colors correspond to the column density through a sphere of radius 100 AU centered on the star particle. The arrows show the direction of the mean in-plane velocity of the disk gas. Bottom - the black circles show the mean angular velocity $\omega$ in the disk as a function of cylindrical radius $R_z$. The red line corresponds to a Keplerian profile normalized using the mass of the star. We have also shown the sink particle accretion zone in green to demarcate the radius at which our sink particle algorithm begins to alter the fluid properties.}
  \label{fig:disk}
\end{figure}

We can also isolate the effect of the magnetic field on the gas morphology by comparing runs BR and HR. There are two main differences. First, without the magnetic field to help support it, the main filament of gas feeding the central protostar has already begun to fragment into self-gravitating, spherical ``beads" by 0.3 free-fall times. These beads have a characteristic size of a few hundred AU, and are therefore well-resolved in our runs. The type of grid-induced filament fragmentation discussed in section 2.2 in the context of isothermal simulations is thus not a concern in runs BR and HR. Second, beginning around the same time, we can see the presence of a dense, $\sim 200$ AU disk around the most massive star in run HR. This disk is centrifugally dominated with a roughly Keplerian velocity profile. We do not see a similar disk in either of our runs with a magnetic field, at least at the 10 AU resolution of the simulations presented here. This is the well-known magnetic braking effect, where at $\mu_{\Phi} = 2$ the field is so efficient at removing angular momentum from the center of the core that it suppresses the formation of a Keplerian disk \citep{allen2003, hennebelle2008, mellon2008}. However, if we repeat run BI with three more levels of refinement so that the maximum resolution is 1.25 AU and the sink accretion radius is 5 AU, we do in fact begin to see a rotationally-dominated disk beginning around $\sim 0.15$ $t_{\rm{ff}}$. By about $\sim 0.2$ $t_{\rm{ff}}$, when the star has reached a mass of about $3.5 M_{\odot}$, the disk has grown to $\sim 40$ AU and developed a Keplerian velocity profile, as shown in Figure \ref{fig:disk}. This would lie entirely within the sink particle accretion zone in our simulation with $10$ AU resolution, so it is not surprising that we do not see it there. While magnetic braking has certainly removed angular momentum from the material accreting onto the disk, allowing it to fall much closer to the central protostar than would be the case without a magnetic field, we do not find that it suppresses the formation of a disk entirely at high resolution. 

Several other researchers have already reported forming disks in MHD simulations of star formation. In a study of magnetic braking in low-mass cores, \cite{hennebelle2009} found that the efficiency of magnetic braking depends on the angle between the initial magnetic field and the core's angular momentum vector, with a 90 degree misalignment lowering the value of $\mu_{\Phi}$ at which disk formation is suppressed by a factor of $2-3$ relative to the aligned case. \cite{santos-lima2012a, santos-lima2012b} studied this problem numerically as well, arguing that the presence of turbulence increases the rate of magnetic diffusion in the inertial range, allowing parcels of gas that have lost magnetic flux to fall onto a disk. \cite{Seifried2012} also found that the presence of turbulent perturbations reduces the efficiency of magnetic braking enough to form a Keplerian disk at $\mu = 2.6$, although they disagree that flux loss is involved. In our disk, we find $\mu$ averaged over a 100 AU sphere around the most massive star has risen to $\sim 20$ by the snapshot displayed in Figure \ref{fig:disk}, although we have not verified that this is due to the mechanism proposed by Santos-Lima et al.

Finally, we briefly mention one more difference between our magnetic and non-magnetic runs: the presence of episodic outflows in runs BI and BR. Around $0.3 t_{\rm{ff}}$, we begin to find material in those runs with radial velocities of $\sim 10$ km s$^{-1}$ away from the primary star. These outflow velocities increase with time, such that by $0.6 t_{\rm{ff}}$ (when the primary has grown to $> 20 M_{\odot}$) they can be as large as $40$ km s$^{-1}$, which is roughly the Keplerian speed at the grid scale. $\sim 10$ km s$^{-1}$ outflows have been observed previously in non-radiative MHD simulations of massive cores \citep[e.g.][]{seifried2011, hennebelle2011}. However, because the outflow launching mechanism is badly under-resolved in our simulations, we shall not discuss outflow properties in detail here.

\subsection{Magnetic Field Structure}

\begin{figure}
  \epsscale{1.15}
  \centering
  \plotone{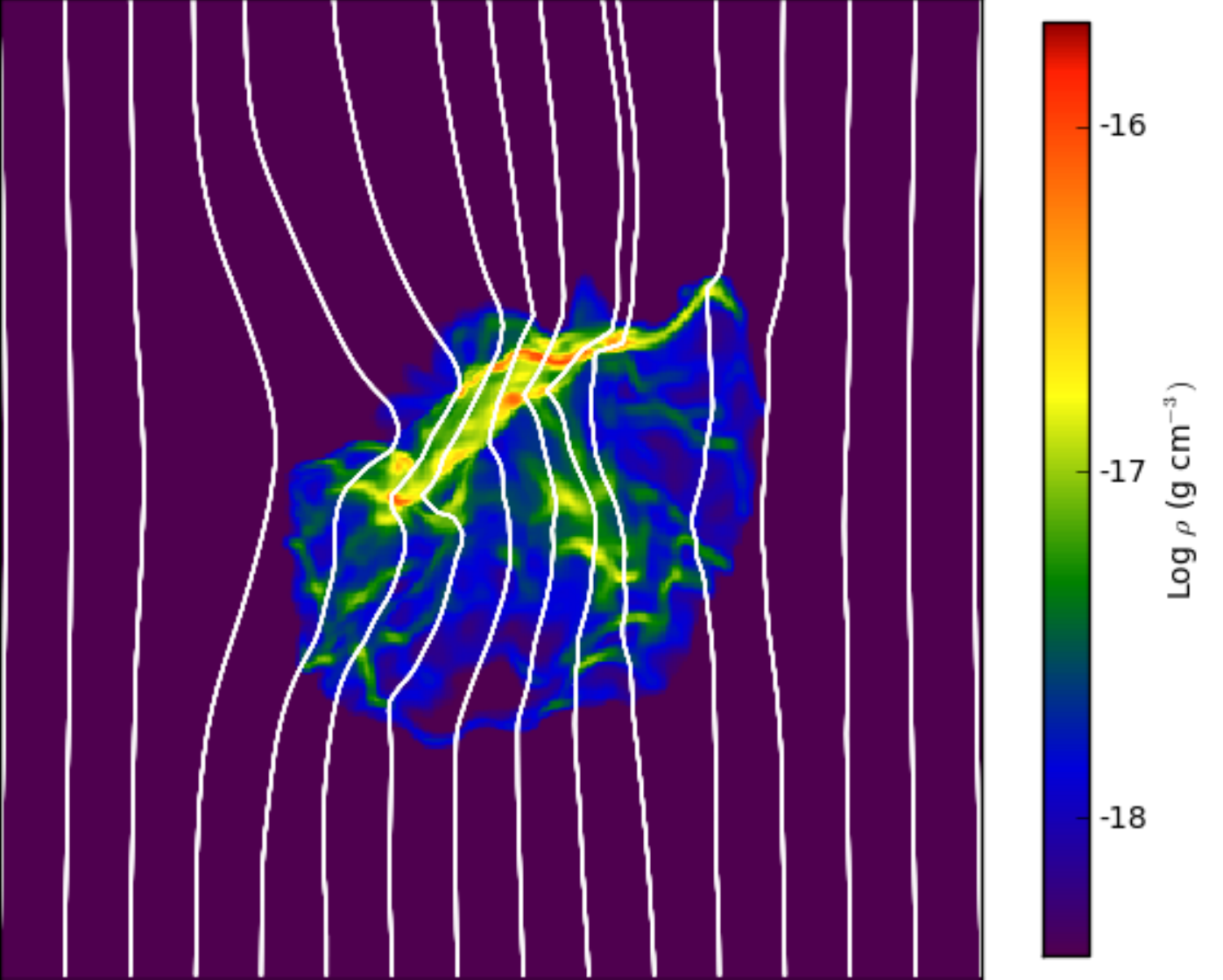}
  \caption{A density slice taken through the center of the computational domain perpendicular to the x-axis at 0.3 free-fall times. The y- and z- components of the magnetic field lines are over-plotted in white with evenly-spaced anchor points along the y-axis.}
  \label{fig:MagneticFieldSlice}
\end{figure}

Although the magnetic field lines are initially oriented in the \it z \rm direction, this is not an equilibrium configuration, and as the simulations proceed they settle into a new, quasi-equilibrium ``hourglass" shape shown in Figure \ref{fig:MagneticFieldSlice}, which resembles the morphology in the dust polarization maps of \cite{girart2009} and \cite{tang2009}. Here, we take a density slice through the center of the domain aligned to be perpendicular to the \it x \rm direction. On top of that slice, we show the planar components (that is, the y- and z- components) of the magnetic field lines. This slice is taken from run BR at 0.3 free-fall times, but the overall shape of the field lines is similar at other times as well, provided enough time has passed for the initial conditions adjust to the new equilibrium. Because the Alfven Mach number of the initial turbulence is $\sim 2$, the lines are able to be bent somewhat by the turbulent perturbations, but this is not a large effect. In the slice shown in Figure \ref{fig:MagneticFieldSlice}, we can see a dense filament in red, with the field lines adjusting so that the magnetic field tends to be perpendicular to the axis of the filament. 

\subsection{Fragmentation and Star Formation}

\begin{figure}
  \epsscale{1.2}
  \centering
  \plotone{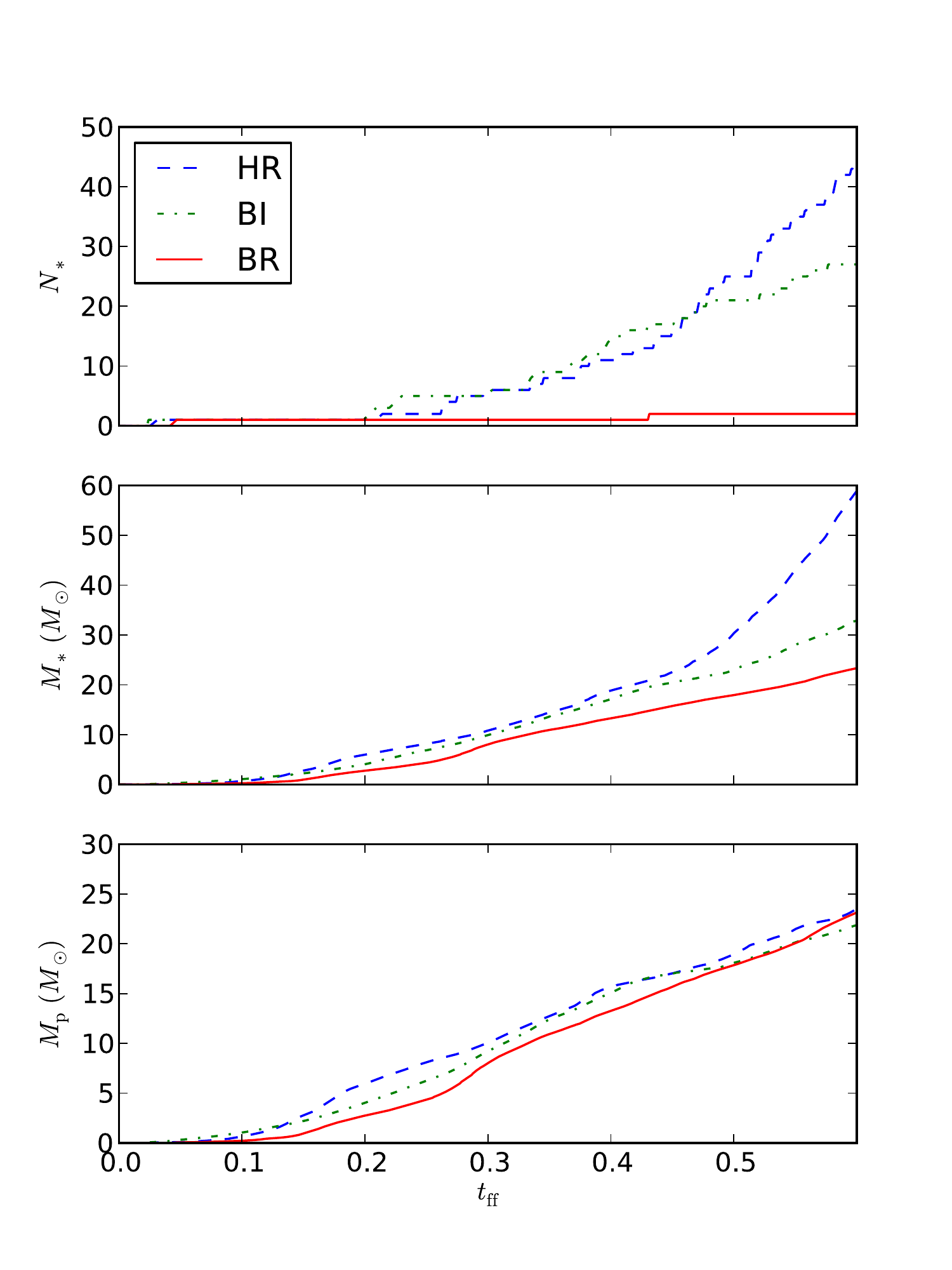}
  \caption{Number of stars $N_{*}$ (top), total stellar mass $M_{*}$ (middle), and mass of the most massive star $M_{\rm{p}}$ (bottom) for all three runs as a function of free-fall time. In this figure and throughout the rest of this paper, we only count a sink particle as a star if it has passed the minimum merger mass of 0.05 $M_{\odot}$, ensuring its permanence as the simulation proceeds.}
  \label{fig:starformationhistory}
\end{figure}

\begin{figure}
  \epsscale{1.2}
  \centering
  \plotone{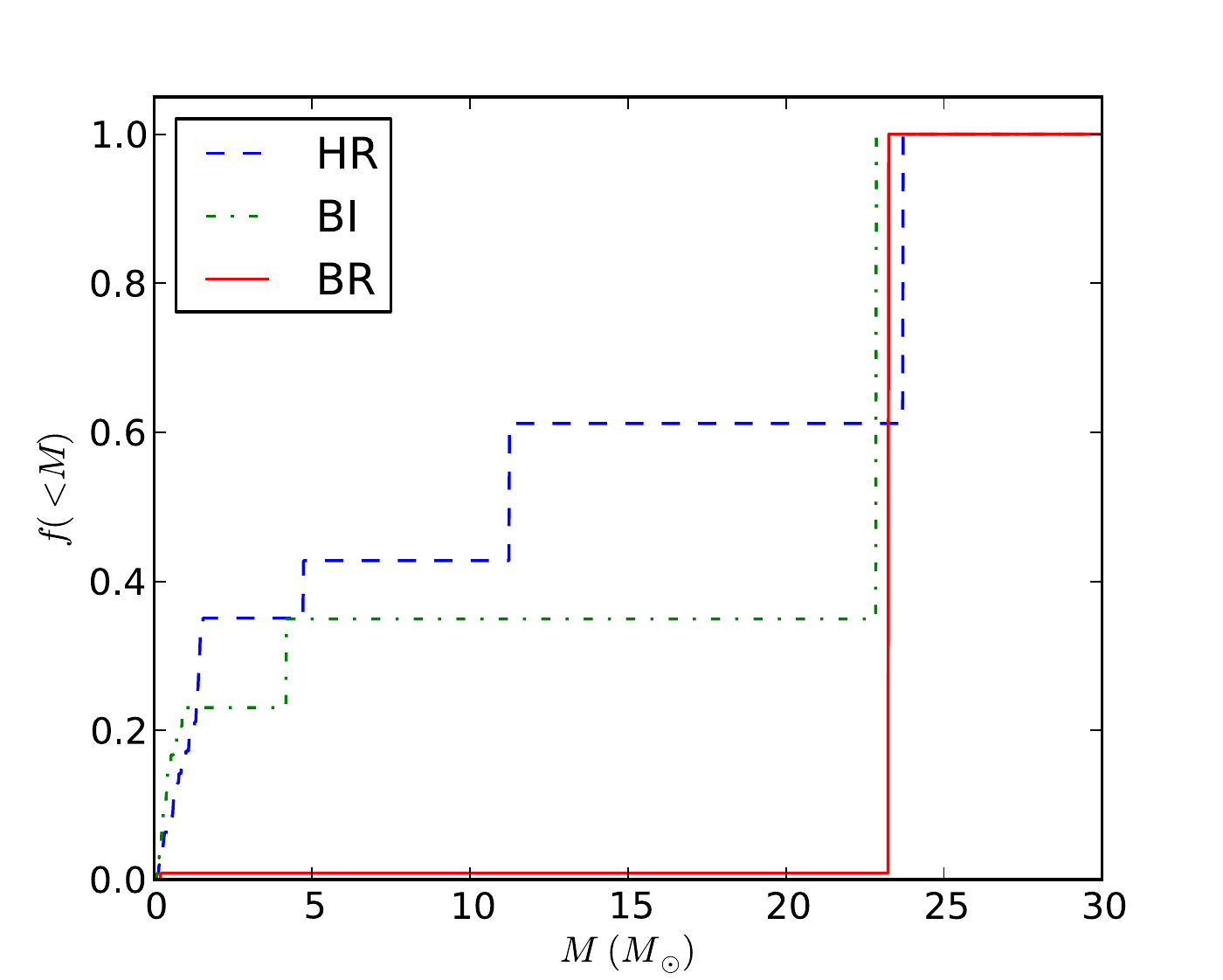}
  \caption{Fraction of total stellar mass that is in stars with mass less than $m$ for all three runs at $t_{ff} = 0.6$.}
  \label{fig:hist}
\end{figure}

The most dramatic difference between the three runs is in the fragmentation. In all three cases, there is a primary with a mass of about $23 M_{\odot}$. In run BR, there is also a secondary star with less than $1 M_{\odot}$ of material. In runs BI and HR, however, the filaments that feed the primary object have fragmented into dozens of stars by the end of 0.6 free-fall times, with typical masses of $0.2 M_{\odot}$ but ranging up to $\sim 11 M{\odot}$. This filament fragmentation takes place beginning around $0.2$ to $0.3 t_{\rm{ff}}$. By $0.5 t_{\rm{ff}}$, these stars have fallen into the central region and undergone significant N-body interactions with each other. After that time, the positions of the sinks in Figures 1 and 2 no longer correspond to the places they were born - many of the sinks have been ejected towards the outer regions of the core. 

We summarize the properties of star particles in all three runs in Figures \ref{fig:starformationhistory} and \ref{fig:hist}. Note that we only count a sink particle as a star once it has passed the minimum merger threshold of $0.05 M_{\odot}.$ Thus, the extra stars in runs BI and HR are not temporary objects that will eventually accrete onto the primary. While the exact value of this threshold is somewhat arbitrary, we point out that in runs BI and HR, there are a few dozen small sink particles that do not meet this threshold by 0.6 free-fall times, while in run BR there are none. Thus, we do not believe that the basic conclusion that fragmentation is dramatically suppressed in run BR compared to the others is sensitive to the exact numerical value of value of the minimum merger mass.

One possible explanation for the difference between runs BR and HR is that extra fragmentation in run HR is due to disk fragmentation that is not present in the other runs because the magnetic field has removed much of the angular momentum from the central region. However, this is not the case. From Figure 2, we can see that run HR has already undergone significant fragmentation in filaments well before the disk has grown large enough to fragment. In fact, most of the stars in run HR form at distances of a few thousand AU or greater from the central star - well outside the disk. Whatever the cause for the difference in fragmentation between the runs with a magnetic field and run HR, it is not due to the presence of a disk in one and not in the others.

As discussed in section 2.2, although much of the fragmentation in run BI is numerical in that it comes from filaments that collapse down to $\rho_{\rm{max}}$, it is still possible to choose $\Delta x_L$ such that the fragment masses are roughly correct. From Equation (\ref{rhomax}), $\rho_{\rm{max}}$ in run BI ranges from $\sim 10^{-14}$ (for $\beta \rightarrow \infty$) to $\sim 10^{-13}$ g cm$^{-3}$ (for $\beta = 0.01$), and so the density at which sink creation occurs in our simulations roughly mimics the density at which molecular gas can no longer cool efficiently. Thus, we expect that the fragmentation in run BI is qualitatively similar to what would happen in massive cores if there was no protostellar feedback: the filaments would fragment a bit after they reached densities of $\sim 10^{-13}$ g cm$^{-3}$, and one would end up with many more fragments than would be formed in the presence of radiative heating.  Furthermore, some of the protostellar properties in run BI do indeed appear to be converged. If we compare both the total mass in stars and the mass of the primary in run BI to the high-resolution version of BI at 0.2 $t_{\rm{ff}}$, we find that they differ by only 6\% and 5\% respectively, over a factor of 8 difference in resolution. Thus, while the number and mass distribution of the fragments in run BI are not converged, quantities that depend mainly on the overall accretion rate do seem to be. 

In addition to the fragmentation, we also find that the magnetic field slows down the overall rate of star formation by about a factor of about 3, consistent with \cite{padoan2011} and \cite{federrath2012}. At 0.6 free-fall times, the total mass in stars in run BR is about $\sim 20 M_{\odot}$, almost all of which is in the primary, compared to over $60 M_{\odot}$ in the run HR. Almost all of the ``extra" star formation in the run HR has gone into stars other than the primary, which contains only $\sim 40$ \% of the total stellar mass at 0.6 free-fall times. The mass of the most massive star, on the other hand, is approximately the same in all three runs, probably because our initial conditions place the same amount of mass in position to quickly collapse towards the center. Beginning at around $0.5 t_{\rm{ff}}$, there is an increase in the rate of star formation in run HR as compared to the others. This increase is associated with the fragmentation of a filament formed in the outer region of the core that by $\sim 0.5 t_{\rm{ff}}$ has begun to form stars, as shown in the bottom panels of Figure 1. The relative timescales here are roughly what one would expect from inside-out collapse given our initial conditions: for a power-law density profile with slope $-1.5$, the ratio of the free-fall time at $0.75 R_c$ to that at $0.25 R_c$ is about a factor of $1.5$, which is approximately the delay we see here. Note that filament fragmentation in the outer regions of the core does not happen in either of the runs with a magnetic field - there, star formation only occurs close to the core center. We will discuss this difference further in section 4.    

We mention here as a caveat that our $10$ AU resolution means that we cannot resolve any binaries closer than $\sim 40$ AU, the accretion radius on one sink particle. Thus, we cannot rule out the possibility that the massive star present in run BR would in fact be massive binary with a separation of $\lesssim 40$ AU if we had higher resolution. However, even if that were the case, the fragmentation would still qualitatively different than in runs HR and BI, where we form dozens of stars with a masses that sample the full IMF.  
 
\subsection{Thermal Structure}

\begin{figure}
  \epsscale{1.1}
  \centering
  \plotone{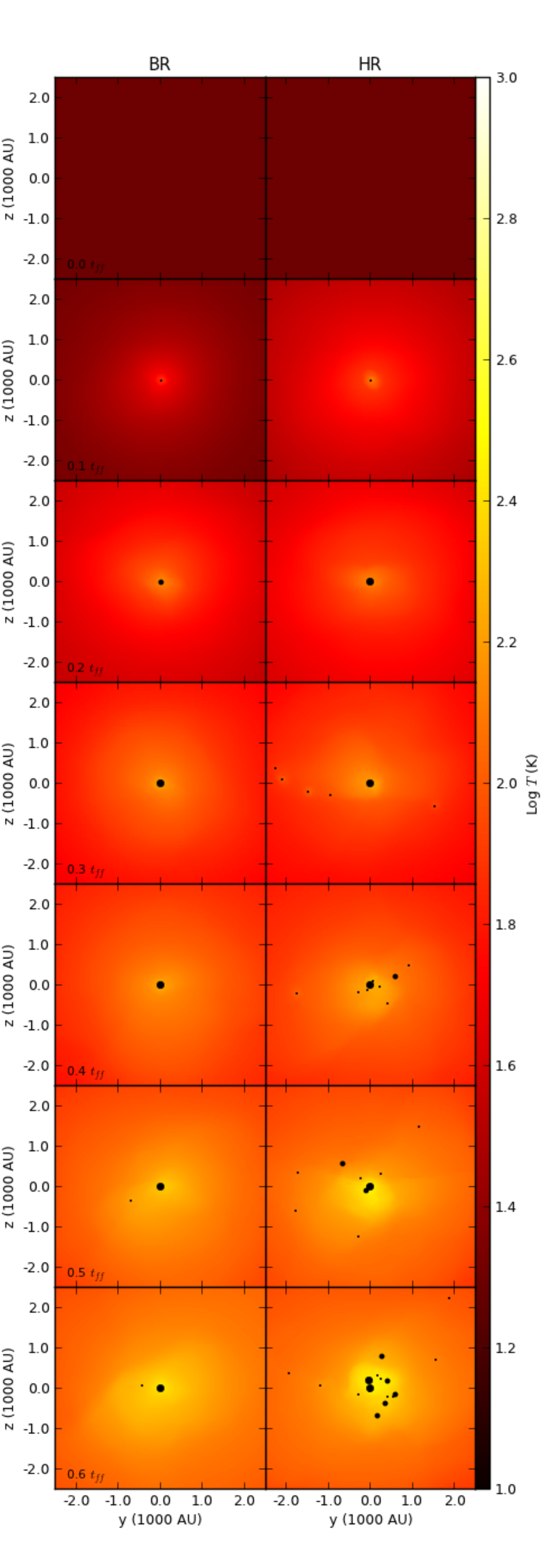}
  \caption{Maps of the average gas temperature, taken at the same times as Figures 1 and 2. The averages were taken along the $x$ direction through a 5000 AU cube around the most massive star. Run BR is on the left and run HR is on the right.}
  \label{fig:TemperatureZoomedIn}
\end{figure}

The difference in fragmentation between runs BI and BR is expected, since it is well-established that radiative feedback in massive cores reduces fragmentation by raising the thermal Jeans mass of the collapsing gas \citep[e.g.][]{krumholz2007b, krumholz2010}. The difference in fragmentation between runs BR and HR, however, is more interesting. One possibility is that protostellar heating is somehow more efficient in the presence of magnetic fields. Figure \ref{fig:TemperatureZoomedIn} shows maps of the average temperature through a 5000 AU cube centered at the most massive star in runs BR and HR. We find that, contrary to this hypothesis, the heating in run HR is either similar to or slightly more widespread than in run BR, because accretion rates are higher in the absence of the field. This is not a dramatic effect, however. The total protostellar luminosity in run BR is typically smaller than that of run HR by only a factor of $\sim 0.7$. The temperatures, which in the optically thin limit scale like $L^{0.25}$, would be lower by only a factor of $\sim 0.9.$ At $0.25 t_{\rm{ff}}$, when the first fragmentation in run HR occurs, the mean $T_g$ in the 5000 AU cube around the primary is 63.3 K in run HR and only 53.7 K in run BR, but despite the higher temperatures the gas in HR fragments while the gas in BR does not. So, the difference in the effectiveness of radiative heating between runs BR and HR cannot be responsible for the difference in fragmentation - it is too small and in the wrong direction.

The difference, then, must be due to the direct support provided by the magnetic field in run BR. To quantify this effect, we define an effective temperature $T_{\rm{eff}}$ by
\begin{equation}
\frac{3}{2} n k_B T_{\rm{eff}} = \frac{3}{2} n k_B T_g + \frac{B^2}{8 \pi},
\end{equation} where $n$ is the number of particles per unit volume. Expressed in terms of $\beta$, we find
\begin{equation}
T_{\rm{eff}} = T_g \left( 1 + \frac{2/3}{\beta}\right).
\end{equation} In other words, $T_{\rm{eff}}$ is the temperature defined in terms of the thermal plus magnetic energy densities instead of just the thermal energy density. Note that this is actually more closely related to our criterion for creating a sink particle than the gas temperature because we have included the magnetic energy in defining the magnetic Jeans number (see the Appendix). While the concept of an effective temperature is clearly an oversimplification - for one, the magnetic field does not resist collapse isotropically the way thermal pressure does - we find that it is helpful in understanding our simulation results.  
 
\begin{figure*}
  \epsscale{1.0}
  \centering
  \plotone{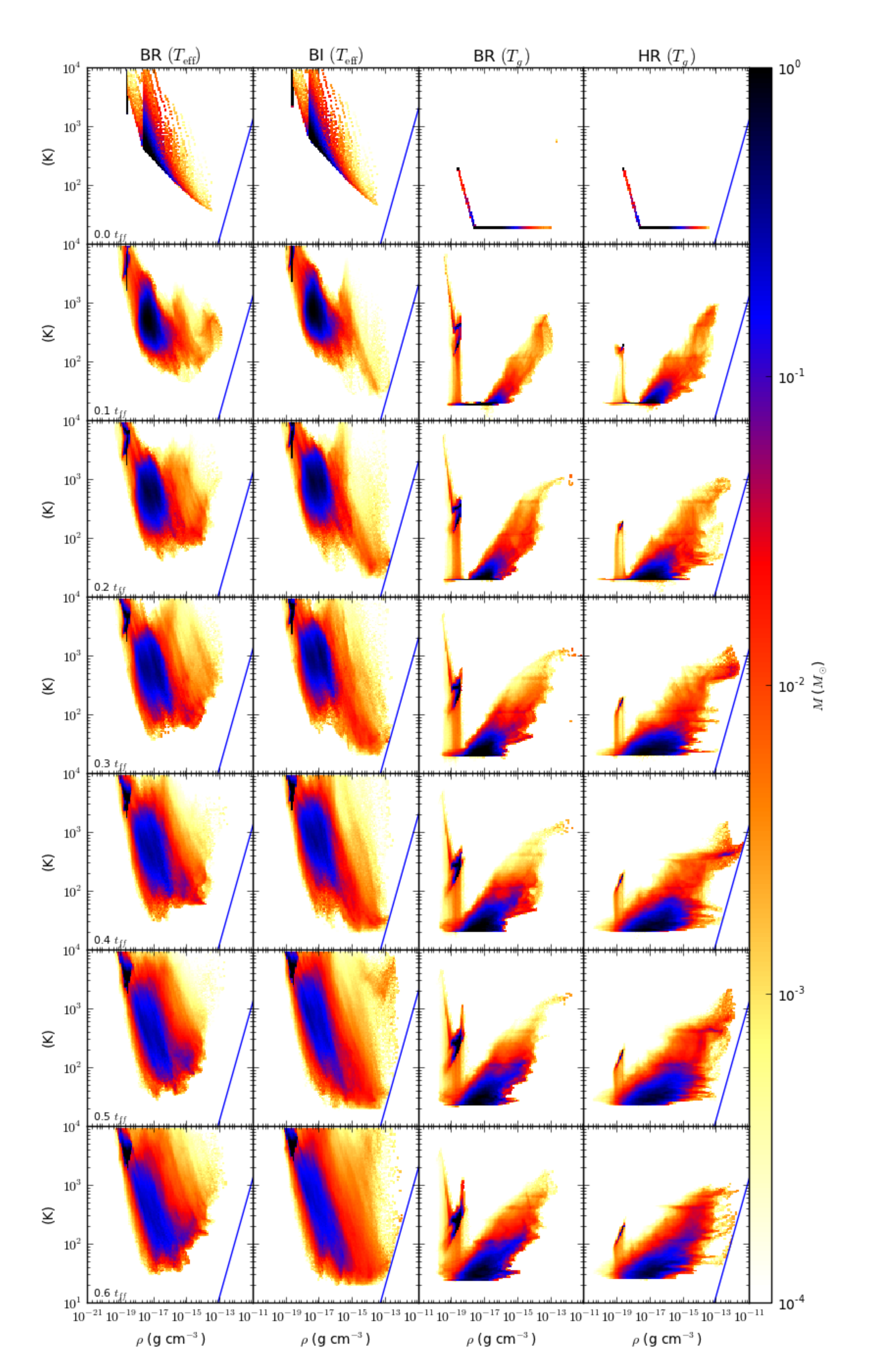}
  \caption{Left two panels - phase diagrams showing the amount of mass in each $\rho - T_{\rm{eff}}$ bin at different times, for runs BR and BI. Right two panels - the same,  but with $\rho - T_g$ bins for runs BR and HR. The snapshots are taken at the same times as the above figures. The islands of low density material at $T_g \sim 10^2$ K  and $T_{\rm{eff}} \sim 10^4$ K correspond to gas in the ambient medium and should be ignored.}
  \label{fig:PhasePlotsEff}
\end{figure*}

 In Figure \ref{fig:PhasePlotsEff}, we summarize the combined temperature and magnetic field structure of the cores in runs BR, BI, and HR. In the two right panels, we plot the total mass in each $\rho - T_g$ bin for runs BR and HR over a series of time snapshots. In the two left panels, we instead use $\rho - T_{\rm{eff}}$ bins for the two runs with magnetic fields. The top row of the figure merely summarizes our initial condition. Although the core temperature starts at precisely 20 K in all three runs, the cylindrically symmetrical magnetic field profile means that there are a range of magnetic field strengths, and thus $T_{\rm{eff}}$ covers a range of values. The blue diagonal lines represent the threshold at which the code lays down a sink particle. Thus, there can be no gas in any of the runs to the right of this line - any cell that exceeds this threshold has some of its gas converted into sink particles until it no longer violates the MHD Truelove criterion. This line is suppressed in the third column, because in the presence of magnetic fields, there is no single density at which sinks are created for a given temperature (see Equation \ref{rhomax}).
 
 We can get a sense of whether star formation is taking place from these plots by looking at whether there is any gas close to crossing this threshold. In run BR, there is hardly any gas close to the densities required for sink formation.  Runs BI and HR, on the other hand, have significant amounts of gas close to that threshold by around $0.2$ to $0.3 t_{\rm{ff}}$. The phase diagram for run HR, in particular, bears a number of ``finger" features that correspond to gas that is all at one $T_g$, but that stretches over a range of densities approaching that required for sink formation. These features are most prominent at $0.3 t_{\rm{ff}}$, but are visible before and after as well. The is precisely the time at which the main filament in run HR has broken up into a number of gravitationally unstable ``beads", which collapse down until they form sink particles. The ``fingers," then, correspond to gas in these beads that is collapsing isothermally, albeit at higher temperatures than the initial 20 K, with the precise value determined by the distance from the bead to the central protostar. This collapse is isothermal because the temperature changes on the evolution timescale of the most massive protostar, which for our problem is $t_{\rm{ff}}$, while the timescale for local gravitational collapse in the bead is must faster.  In contrast, we do not see this behavior in run BR, because a combination of magnetic and thermal support has rendered the main filament in that run stable against gravitational collapse at a density much higher than the sink creation value. 
 
 In one sense, Figure \ref{fig:PhasePlotsEff} restates what we already know - there is much fragmentation in runs BI and HR and hardly any in run BR. However, this plot can also help us untangle the effect of the magnetic field and the radiative feedback by telling us in which regimes each effect is more important. By comparing runs BR and BI, for instance, we can see that the primary effect of the radiation is to heat up the relatively dense regions in the core - i.e. to move material greater than about $1 \times 10^{-15}$ g cm$^{-3}$ up in the plot and away from the sink formation threshold. Alternatively, the slope of the $(\rho - T_{\rm{eff}}$) phase diagrams for runs BR and BI show that the magnetic field is most effective at raising $T_{\rm{eff}}$ at low density. Hence, we can begin understand that the reason the combination of the $\vecB$ field and the radiative feedback is more effective at suppressing fragmentation than either considered in isolation is that they are effective in different regions, with the magnetic field mostly helping to support (or, at least, to slow the collapse of) material in the diffuse, outer parts of the core, and with radiation most effective in the dense material that is close to the central protostar. 
 
 \section{Discussion}
 
 \subsection{Why do Magnetic Fields and Radiation Suppress Fragmentation?} 
We would like to understand the suppression of fragmentation in run BR in terms of the mass-to-flux ratio $\mu_{\Phi}$. The average $\mu_{\Phi}$ for the entire core is 2, but, because the core is centrally concentrated, it is greater through flux tubes passing near the center and lower through flux tubes passing through the diffuse, outer regions. It is illustrative to do the following analysis on our initial conditions: take the initial spherical region and exclude a cylindrical region of radius $R_z$ concentric with the sphere and extending through the entire domain. Then, compute $\mu_{\Phi,z}$, the mass-to-flux ratio in the remaining region. $\mu_{\Phi,z}$ is 2 when $R_z = 0$ and monotonically drops to 0 when $R_z = R_c$ How quickly $\mu_{\Phi,z}$ drops off with $R_z$ will give us a rough estimate of where we can expect the core to be subject to fragmentation. We find that, by a radius of $R_z \approx 0.73$ $R_c$, $\mu_{\Phi,z}$ has dropped below 1, meaning that the region external to that cylindrical radius (corresponding to approximately 32\% of the core volume and 19\% of the mass) should be fairly well-supported against collapse. Furthermore, the point at which $\mu_{\Phi,z}$ has dropped to 1.5 is at only $0.44$ $R_c$, meaning that $\sim$72\% of the core volume and $\sim$53\% has a mass-to-flux ratio below that value. While structures with a mass-to-flux ratio of 1.5 are supercritical and should collapse, they will still collapse more slowly than in the absence of the magnetic field, giving the radiative feedback more time act. This effect is not dramatic; the effect of the magnetic pressure force in the virial theorem is to dilute gravity along the field lines by a factor of $(1 - \mu_{\Phi}^{-2})$ \citep{shu1997}, so that structures that are supercritical by a factor of 1.5 collapse approximately half as quickly as structures with $\mu_{\Phi}$ of infinity, and even the core as a whole collapses about 75\% as fast at $\mu_{\Phi} = 2$. 

\begin{figure}
  \epsscale{1.2}
  \centering
  \plotone{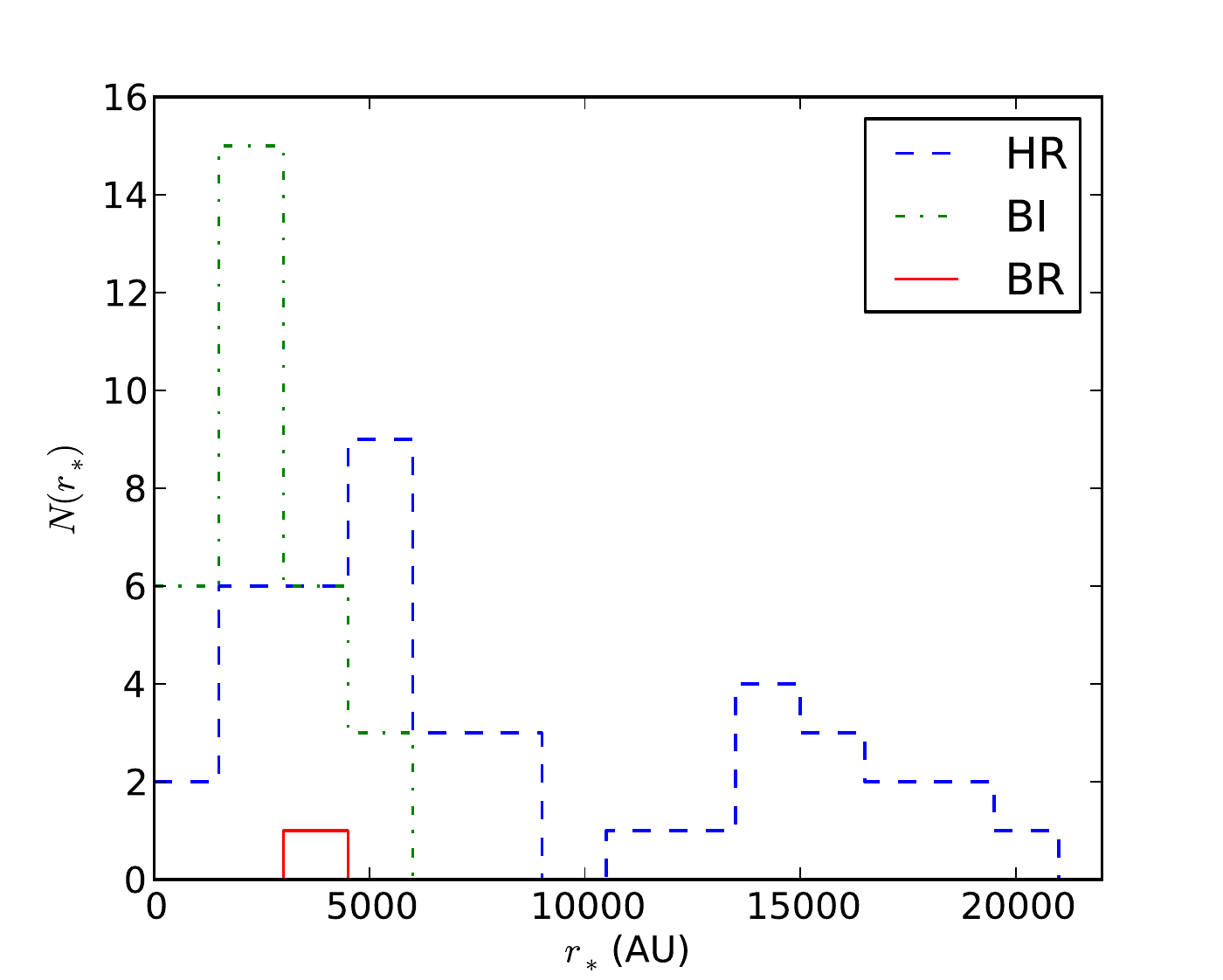}
  \caption{Histograms of star formation distance $r_*$ for runs BI, HR, and BR. Here, $r_*$ is the distance each star was from the most massive star when it formed, computed for every star that forms over the entire history of each simulation. In run BR, there is only one secondary fragment, which forms a distance of $r_* \approx 3600$ AU from the primary.}
  \label{fig:stardist}
\end{figure} 

Thus, even inside a supercritical core, the magnetic field can slow collapse in specific sub-regions with mass-to-flux ratios below unity, or even halt it altogether, and this mechanism is more effective at suppressing fragmentation in the outer regions of the core. We can see this effect operating in the bottom row of Figure 1 - while most of the star formation in run HR takes places towards the central region of the core, by 0.6 free-fall times we have begun to see signs of fragmentation of a filament in the outer regions as well. This does not take place in either of the runs with magnetic fields, including the one with no radiation, so this cannot be a radiative effect. Rather, it is due to the ability of the magnetic field to effectively suppress fragmentation in the outer regions of the core. To quantify this, we plot in Figure \ref{fig:stardist} the distribution of $r_*$, which measures how far away each star was from the central massive star at the time it formed. We compute this quantity for every star (other than the first) that forms over the history of each simulation. In run BI, most of the fragmentation takes place at distance of $\sim 2,000$ AU, and there are no stars that form at a distance greater than $\sim 5,000$ AU from the central object. In run HR, however, the most likely value of $r$ is roughly 5000 AU, while a significant fraction (about 1/3) of the stars form at distances of 10,000 AU or greater. Star formation at such large radii is completely suppressed by the magnetic field, even without heating effects.

The following picture thus emerges: magnetic fields work to suppress fragmentation in the outer regions of the centrally-concentrated cores, either by slowing it down or halting it altogether. If they halt it altogether, then fragmentation is confined to the central region, where radiative heating is most effective. If magnetic fields merely slow fragmentation at large radii, then they still allow radiative heating more time to ``win" by heating up filaments to the point at which they are too warm to collapse further. The combined result is that the magnetic field and the radiation are together far more effective at suppressing fragmentation than either process in isolation. Our work suggests that typical massive cores, which are centrally concentrated and have $\mu_{\Phi} \sim 2$, do not fragment strongly, as one would expect from the correspondence between the core and initial stellar mass functions. 

 Finally, while our simulations are based on ideal MHD, we do not expect that non-ideal effects will dramatically alter our conclusions. Ohmic dissipation, ambipolar diffusion, and reconnection diffusion \citep{santos-lima2010, lazarian2011} are all capable of increasing $\mu_{\Phi}$, but they also all most important at high densities and/or regions where the magnetic field lines are most bent. Those are precisely the regions where the magnetic field is $\emph{least}$ important for the suppression of fragmentation, because they are all primarily associated with the inner region of the core where radiative heating is most effective.  
 
 \subsection{Comparison to Commer{\c c}on et. al.}
\cite{commercon2011} have also performed a set of radiation-magnetohydrodynamic simulations of massive core collapse using similar initial conditions to the ones considered here. They also found a synergistic effect between the radiative heating and the magnetic field, where the two effects in tandem lead to much less fragmentation than either considered in isolation.  However, their work differs from our own in a few key respects. Most significantly, they have focused on the early stages of collapse, up to just past the point at which the first hydrostatic core forms, while we have focused on what happens to the remainder of the core after the first protostar has undergone second collapse. Thus, while we have both identified mechanisms by which a combination of radiative and magnetic effects suppress fragmentation, these mechanisms have different underlying causes and manifest themselves at very different times. 

This difference in emphasis stems from our different recipes for representing protostellar feedback. In this work, we use sink particles to represent material that has collapsed to densities higher than we can follow on the grid, and compute a luminosity $L_i$ for each particle according to a sub-grid model. \cite{commercon2011}, on the other hand, do not use sink particles, but instead employ higher resolution ($2.16$ AU), which allows them to follow the formation of a hydrostatic core. The feedback from the accretion shock on to this core can then be computed on the grid. They find the radius at which this shock releases its energy depends on the core's magnetic field, since strong magnetic braking allows smaller first cores to form. 

The approach of \cite{commercon2011} has the advantage of self-consistency. Furthermore, it is probably more accurate than our technique at representing radiation from particles before the second core has formed. In fact, we do not include any radiative feedback until the sink particle mass exceeds $0.01 M_{\odot}$, and second collapse generally takes place at a few times that value, so except for a very short period of time, we ignore this radiation altogether. Also, because their resolution is higher and they do not have to handle sink particle mergers, they can resolve binaries that we cannot. However, their method has the downside that it cannot model the effect of the much larger (by a factor of $\sim 100$) accretion luminosities that occur after second collapse. Additionally, the lack of sink particles severely limits the integration time for \cite{commercon2011}, since they cannot follow collapse past the point where they fail to resolve the Jeans density. This limited them to running for only a few percent of a free-fall time after the first hydrostatic object formed. In contrast, with our initial conditions the first core forms almost immediately, and we find that most fragmentation does not occur until around 20\% to 30\% percent of a free-fall time past that point. In a sense, our simulations pick up where those of \cite{commercon2011} left off, in that our simulations begin with a centrally concentrated core with one protostar that very quickly undergoes second collapse. 

Sub-grid luminosity models have problems of their own related to unresolved binarity, as pointed out by \cite{bate2012}. As mentioned in section 3.3, our use of sink particles means that we cannot resolve any binaries closer than 40 AU, and we cannot rule out the possibility that the central massive star in our simulations in fact represents an unresolved binary. However, for accretion luminosity-dominated stars, it matters little whether a sink particle represents a single star or a binary too tight to be resolved, because the energy released per unit mass accreted onto low-mass protostars is nearly independent of the stars' masses \citep{krumholz2011}. On the other hand, stars' internal luminosity scales with mass as roughly $M^{3.5}$, meaning that in the worst case where a sink particle should in fact represent an equal mass binary, the internal luminosity is overestimated by a factor of $2^{2.5} = 5.7$. While this is a potential concern, the internal luminosity does not become comparable to the accretion luminosity in our calculations until about $0.3 t_{\rm{ff}}$, and by that time there are already clear differences in the fragmentation between runs HR and BR. Moreover, the alternative of not including a sub-grid luminosity model, as in \cite{bate2012}, is far worse. Without such a model one omits both the accretion luminosity onto the stellar surface and the larger internal luminosity, and the resulting error is many orders of magnitude.

Finally, we mention one last difference between our work and \cite{commercon2011}: their initial conditions contained much less kinetic energy than our own, with $\alpha_{\rm{vir}} = 0.2$ versus $\alpha_{\rm{vir}} = 2.3$. This could explain why we see a small disk in our high resolution run, and \cite{commercon2011} do not. If their $\beta_{\rm{rot}} \approx 0.02 \alpha_{\rm{vir}}$, as implied by \cite{BB2000}, then they would have $\beta_{\rm{rot}} \approx 0.004$, smaller than our own by a factor of 3. On the other hand, \cite{Seifried2012} had $\beta = 0.04$, higher than ours by a factor of 4 again. Thus, the sequence of disk sizes seen in our papers, ranging from $\sim 100$ AU \citep{Seifried2012} to $\sim 40$ AU (us) to unresolved \citep{commercon2011}, could simply be a consequence of different amounts of angular momentum in the cores.  It is possible, however, that if \cite{commercon2011} extended their simulation to later times, that they too would begin to resolve a disk in their $\mu_{\Phi} = 2$ run. We conjecture that this disk would be smaller than $\sim 40$ AU in radius.  

\subsection{Where is Fragmentation Suppressed?}
An interesting question is: in what range of the $\Sigma - \mu_{\Phi} $ parameter space is fragmentation weak? As discussed in \cite{crutcher2010}, although the average molecular cloud core is marginally magnetically supercritical, it by no means follows that there are no cores with weak magnetic fields. The Bayesian analysis presented in that paper suggests the distribution of field strengths is quite flat, such that there may be many cores where the field is significantly weaker than the ones discussed here. In fact, for cores like ours with a mean density of $n_{\rm{H}} = 2.4 \cdot 10^{6}$ cm$^{-3}$, their result suggests that the total magnetic field strength should be evenly distributed between $\approx 0.0$ mG and $\approx 3.4$ mG, roughly twice the value considered here. This implies that about 25\% of cores like the ones in this paper would have values of $\mu_{\Phi}$ of 4 or greater. Our work suggests that there should be a tendency towards greater fragmentation in massive cores with such weak magnetic fields, with such cores being more likely to form clusters rather than isolated massive stars or binaries. A recent set of millimeter observations \citep{palau2013} studied the fragmentation of 18 massive cores with $\lesssim 1000$ AU resolution, and found that $\sim 30 \%$ showed no signs of fragmentation, while $50 \%$ did. They propose that variation in the magnetic field strength may be responsible for the determining the fragmentation, but confirmation of this view will have to wait for follow-up observations of the field.  

Furthermore, the recent observations of \cite{butler2012} found a typical massive core surface density of  $\sim 0.1$ g cm$^{-2}$, over a factor of 10 lower than the $2$ g cm$^{-2}$ cores considered here. These cores are below the surface density threshold for massive star formation $\sim 1$ g cm$^{-2}$ identified in \cite{krumholz2008}, which ignored magnetic fields. Could magnetic fields play a role in lowering the threshold for massive star formation? We plan to address these questions in future work. 
   
 \section{Conclusions}
 
 We have presented a set of 3D, R-MHD simulations of the collapse of isolated, magnetized, massive molecular cloud cores that attempt to isolate the effects of the magnetic field and of the radiative feedback on core fragmentation. We find that the magnetic field and protostellar radiation can combine to largely suppress fragmentation throughout the core, so that the simulation that includes both magnetic fields and radiation results in only a single binary star system, while the runs that exclude either effect are subject to far more fragmentation. The explanation for this behavior is that magnetic fields and radiative heating are effective in different regimes, so that that each effect influences gas that the other misses. We find that massive cores with typical magnetic field strengths likely collapse to form single star systems, as suggested by the observed relationship between the CMF and IMF. We have also reproduced the result found by other researchers that Keplerian disks can form in the presence of magnetic fields with $\mu_{\Phi} \sim 2$, provided that turbulence, which results in a misalignment between the magnetic field and angular momentum vectors, is present. 
 
 \acknowledgements
A.T.M. wishes to thank Pak-Shing Li, Louis Howell, Christoph Federrath, and Bo Zhao for helpful discussions, and the anonymous referee for constructive comments that improved the paper. Support for this work was provided by NASA through ATP grant NNG06-GH96G (R.I.K., M.R.K. and C.F.M.) and a Chandra Space Telescope grant (M.R.K.); the NSF through grants AST-0908553 and NSF12-11729 (A.T.M., R.I.K. and C.F.M.) and grant CAREER-0955300 (M.R.K.); an Alfred P. Sloan Fellowship (M.R.K); and the US Department of Energy at the Lawrence Livermore National Laboratory under contract DE-AC52-07NA27344 (A.J.C. and R.I.K.) and grant LLNL-B569409 (A.T.M.). Supercomputing support was provided by NASA through a grant from the ATFP. We have used the YT toolkit \citep{yt} for data analysis and plotting. 

\begin{appendix}
 \section{The MHD Truelove Condition}

Simulations of isothermal, self-gravitating systems are subject to artificial
fragmentation unless the Jeans length,
\beq
\lj\equiv \left(\frac{\pi c_s^2}{G\rho}\right)^{1/2},
\eeq
is resolved by a sufficiently large number of grid cells,
\beq
\jth\equiv\frac{\Delta x}{\lj}<\jthmax,
\eeq
where $\Delta x$ is the width of a grid cell \citep{truelove97}. 
We have added the subscript ``th" to the Jeans number $J$ to indicate
that it is for the purely thermal case, in which there is no magnetic field. For the case
they studied, \cite{truelove97} found that $\jthmax=0.25$ was adequate
to suppress artificial fragmentation, but in general it is a problem-dependent quantity.
For AMR simulations, the Truelove condition is
one of the criteria used to increase the refinement; for
simulations in which sink particles are used as a sub-grid model
for protostars, the Truelove condition is often
used to determine when sink particles should be introduced---i.e., whenever
the density exceeds
\beq
\rho_{\max}=\frac{\pi \jthmax^2 c_s^2}{G\Delta x^2}.
\label{eq:rhomaxth}
\eeq

Magnetic fields suppress fragmentation and therefore should allow one to
defer refinement or the introduction of sink particles to higher densities. \cite{federrath2010} included the effects of the magnetic field in their refinement criteria, but it was used as a supplement to the thermal Truelove
criterion, not as a replacement. For introducing sink particles, they did require that the total energy of a control volume be negative.

To generalize the Truelove condition to include magnetic fields, we begin
with the expression for the maximum mass of an isothermal, magnetized 
cloud derived by \cite{mouschovias76}, as generalized by \cite{tomisaka88},
\beq
\mcr=1.18\mbe\left[1-\left(\frac{M_\Phi}{\mcr}\right)^2\right]^{-3/2},
\label{eq:mcr1}
\eeq
where 
\beq
\mbe=1.18\left[\frac{c_s^3}{(G^3\rho)^{1/2}}\right]
\eeq
is the Bonnor-Ebert mass and
\beq
M_\Phi\simeq\frac{\Phi}{2\pi G^{1/2}}
\eeq
is the magnetic critical mass. (\cite{tomisaka88} found that a factor 0.17
fit their numerical results better than $1/(2\pi)$, but we adopt the latter for
simplicity.) Using the alternative form for the magnetic
critical mass, $M_B$, which is defined by
\beq
\frac{M_B}{\mcr}=\left(\frac{M_\Phi}{\mcr}\right)^3,
\eeq
Equation (\ref{eq:mcr1}) can be expressed as 
\beq
\mcr=\left[1.12\mbe^{2/3}+M_B^{2/3}\right]^{3/2}
\eeq
(\cite{bertoldi92}, who wrote $M_{\rm J}$ for the Bonnor-Ebert mass).
Evaluation of $M_B$ gives $M_B/\mbe=0.76\beta^{-3/2}$. We define the
critical radius by \citep{mouschovias76}
\beq
\mcr=\frac 43 \pi\rho\rcr^3,
\eeq
and then obtain
\beq
\rcr=0.39\lj\left(1+\frac{0.74}{\beta}\right)^{1/2}.
\eeq
We denote the critical radius in the absence of a magnetic field ($\beta\rightarrow\infty$)
as $\rcrth$. As expected, we see that $2\rcrth\simeq \lj$.

In the non-magnetic case, the Jeans number is $\jth=\Delta x/\lj\propto \Delta x/\rcrth$.
We then generalize the Jeans number to the MHD case by writing
\beq
\frac{J}{\jth}=\frac{\rcrth}{\rcr},
\eeq
so that
\beq
\jth=J\left(1+\frac{0.74}{\beta}\right)^{1/2}.
\eeq
Since the same relation applies to the maximum Jeans numbers, the MHD Truelove
condition follows from Equation (\ref{eq:rhomaxth}):
\beq
\rho_{\rm max}=\frac{\pi \jmax^2 c_s^2}{G\Delta x^2}\left(1+\frac{0.74}{\beta}\right).
\label{eq:rhomax}
\eeq
We note that if one expresses the Jeans length in terms of the energy density, $u=\frac 32 \rho c_s^2$,
as $\lj=(2\pi u/3G \rho^2)^{1/2}$ and then adds the magnetic energy density into $u$, one obtains the same
result as in Equation (\ref{eq:rhomax}) except that the factor 0.74 is replaced by $\frac 23$, a negligible
difference. Our result is thus very similar to the approach advocated by \cite{federrath2010}.
The advantage of the present derivation is that it is directly tied to the maximum stable mass.

Magnetic fields can halt collapse perpendicular to the field, but they have no effect on
gravitational instability parallel to the field \citep{chandra61}. Gas that collapses along
the field lines has a thickness 
\beq
H=\frac{\Sigma}{\rho_0}=\frac{\surd 2}{\pi}\;\lj,
\eeq
where $\Sigma$ is the surface density, so that
\beq
H=\left(\frac{0.45}{\jth}\right)\Delta x.
\label{eq:H}
\eeq
In fact, a self-gravitating sheet cannot become thinner than $2\Delta x$, since a single layer
of cells cannot exert a vertical gravitational force inside the layer. Hence, if it is important
to follow the internal dynamics of gravitationally stable sheets, one should maintain
$2\Delta x<H$, corresponding to $\jth\la 0.25$ from Equation (\ref{eq:H}). This criterion
does not apply to gravitationally unstable sheets ($J>\jmax$), since they will either be refined
or replaced by sink particles.

\begin{figure}
  \centering
  \epsscale{0.9}
    \plotone{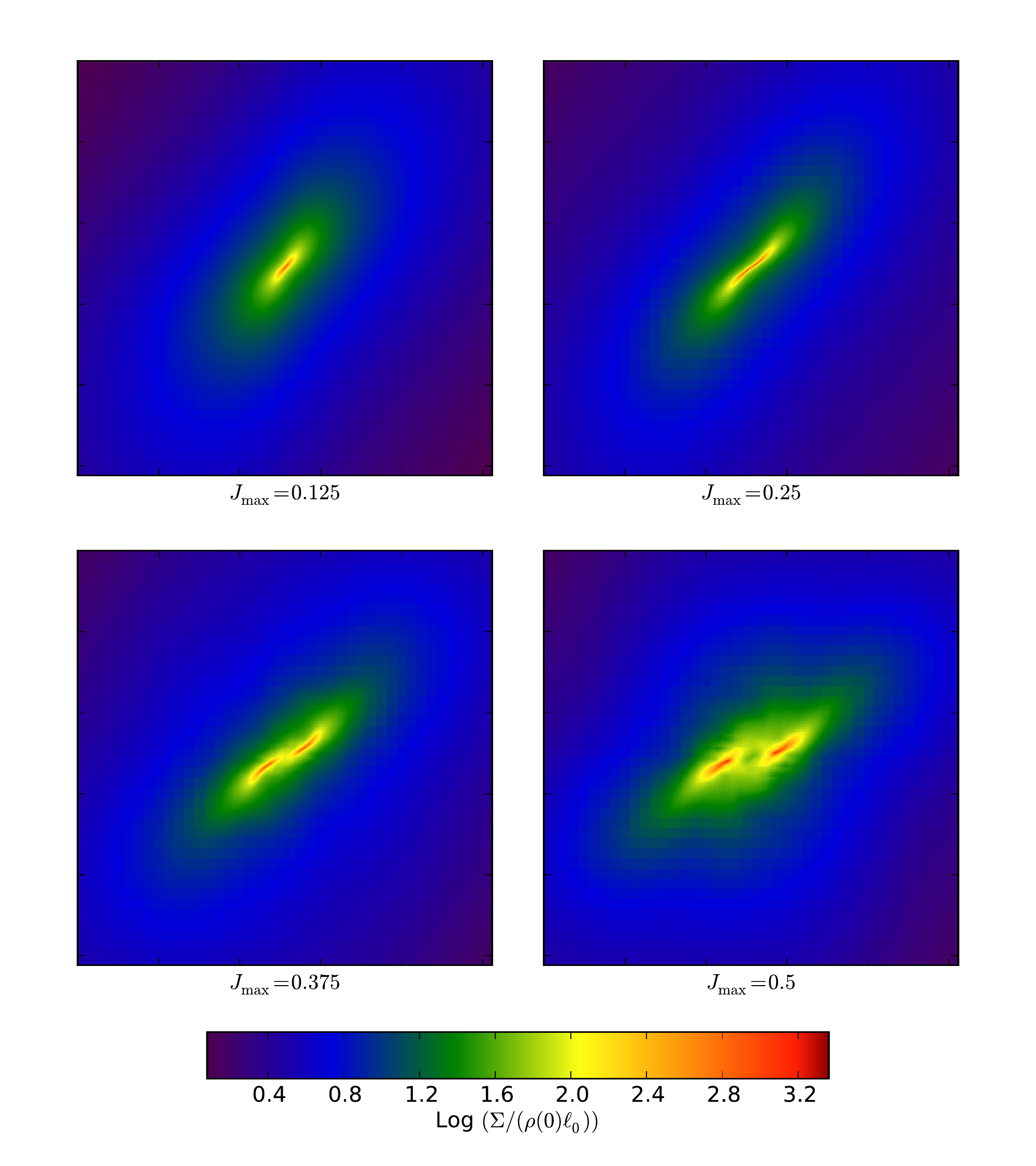}
  \caption{Logarithm of the column density (normalized to $\rho(0) \hspace{3 pt} \ell_0$) through the filament in each run at the point at which the maximum density has reached $2 \times 10^{7}$ times the initial value. Top Left - $\jmax = 0.125$. Top Right - $\jmax = 0.25$. Bottom Left - $\jmax = 0.375$. Bottom Right - $\jmax = 0.5$. The top panels have only one filament, while the bottom two show clear signs of artificial fragmentation. Each image shows a region of size $\approx 0.07 \hspace{3 pt} \ell_0.$}
  \label{fig:MagneticTruelove}
 \end{figure}

To test the MHD Truelove criterion, we carry out the same test used by \cite{truelove97}, 
but with the addition of an initially uniform magnetic field. We begin with a cubic, periodic box of size $\ell_0$ filled with isothermal gas that has a spherically symmetric, Gaussian density profile 
\beq
\rho(r) = \rho(0) \exp{ \left[-\left( \frac{r}{r_1} \right)^2 \right]}.
\eeq Here, $\rho(0)$ is the central density and $r_1$ is a characteristic fall-off radius, which have taken to be $ 0.48 \hspace{3 pt} \ell_0.$ To this background density we also added a $m=2$ azimuthal density perturbation with an amplitude of 10\%. We report our simulation results in units that have been normalized by $\rho(0)$, $\ell_0$, and the central-density free-fall time, given by
\beq
t_{\rm{ff}} = \sqrt{\frac{3 \pi}{32 G \rho(0)}}.
\eeq We have also set the core in initial rotation about the \it z \rm axis with an angular velocity $\omega$ such that $\omega {t_{\rm{ff}}} = 0.53$. This value was chosen so that the core would make approximately 1 rotation in 12 free-fall times, as in the original \cite{boss79} version of this problem. Finally, we have imposed an initially uniform magnetic field field pointing in the \it z \rm direction, with a magnitude such that $M / M_\Phi \approx 2.25$. This field is strong enough to alter the morphology of the collapse, but not strong enough to halt it completely. This value of the mass-to-flux ratio is also comparable to that observed in real star-forming regions.  The initial plasma beta $\beta_0$ is $\approx 1.57$ at the center of the domain and as low as $0.19$ at the edges.

As in the non-magnetic case, the gas first collapses into a sheet and then into a filament, where the field is normal to the filament. The plasma beta in the sheet before it begins to collapse is of order unity. Since the magnetic energy and the gravitational energy are both independent of the radius of the filament, an isothermal filament with a supercritical mass-to-flux ratio will collapse indefinitely \citep{inutsuka2001}. To see this, note that the gravitational energy per unit length is $-Gm_\ell^2$ \citep{fiege2000}, where $m_\ell$ is the mass per unit length, whereas the magnetic energy per unit length is of order $\pi r^2 B^2\propto \Phi_\ell^2$, where $\Phi_\ell$ is the magnetic flux per unit length. (The exact expression for the magnetic energy depends on the structure of the field inside the filament.) Thus, the ratio of the two forces depends only on the mass-to-flux ratio per unit length, which is constant in ideal MHD. Note, however, that the scaling of the magnetic field is only valid if the field is perpendicular to the filament axis. 

We have run four versions of this problem, each time using the AMR capabilities of our code to impose a different $\jmax$. The results are shown in Figure \ref{fig:MagneticTruelove}. Because the free-fall time is a function of density, more or less well-resolved simulations of this problem will not be at the same stage of development at the same simulation time. We instead compare the runs at the point where they have all reached approximately the same maximum density of $2 \times 10^{7} \hspace{3 pt} \rho(0)$. The simulation times at which the peak density reaches this value range from about $3.79 \hspace{3 pt} t_{\rm{ff}}$ in the best-resolved case to $4.02 \hspace{3 pt} t_{\rm{ff}}$ in the worst. We find that, as in the pure hydro version of this problem, $\jmax = 1/4$ is sufficient to halt the onset of artificial fragmentation. The maximum thermal Jeans number, on the other hand, is larger then $\jmax$ by a factor of $\approx 2$ in these runs. Refining on the less stringent magnetic Jeans number seems sufficient to accurately follow the collapse of magnetized gas for this problem, although other physical processes, such as B-field amplification via dynamo action, may require a higher resolution \citep{federrath2011}. 

\end{appendix}

\bibliographystyle{apj}
\bibliography{bibliography}	
 
\end{document}